\documentclass[12pt]{article}

\usepackage{ytableau}
\usepackage[vcentermath]{youngtab}
\usepackage{yfonts}
\usepackage{hyperref}
\usepackage{epsfig}
\usepackage{graphicx}
\usepackage{amsfonts} 
\usepackage{amsmath}
\usepackage{color} 
\usepackage{pstool}
\usepackage[all]{xy}
\usepackage{mathtools}
\usepackage{xspace}

\usepackage[numbers,sort&compress]{natbib}

\graphicspath{ {./fig/} }

\newcommand{\so}{\mathfrak{so}}

\def\beq{\begin{equation}} 
\def\eeq{\end{equation}} 
\newcommand{\be}{\begin{equation}}
\newcommand{\ee}{\end{equation}} 
\newcommand{\ba}{\begin{eqnarray}}
\newcommand{\ea}{\end{eqnarray}}

\newcommand{\Acal}{{\mathcal A}}

\newcommand{\Dcal}{{\mathcal D}}

\newcommand{\Fcal}{{\mathcal F}}

\newcommand{\Hcal}{{\mathcal H}}

\newcommand{\Ocal}{{\mathcal O}}

\newcommand{\Tcal}{{\mathcal T}}

\newcommand{\Vcal}{{\mathcal V}}

\newcommand{\calW}{\mathcal{W}}

\let\a=\alpha \let\b=\beta \let\g=\gamma \let\d=\delta \let\e=\epsilon
    
\let\l=\lambda \let\m=\mu \let\n=\nu  \let\r=\rho
\let\s=\sigma \let\t=\tau    
\let\w=\omega  \let\D=\Delta

\def\subalg{\mathfrak{p}\xspace}

\def\IC{\textrm{I}\xspace}
\def\IIA{\textrm{II}\xspace}
\def\IIID{\textrm{III}\xspace}
\def\IVB{\textrm{IV}\xspace}
\def\VE{\textrm{V}\xspace}

\newcommand{\TypeI}{{\textrm{\bf Type \IC}}}
\newcommand{\TypeII}{{\textrm{\bf Type \IIA}}}
\newcommand{\TypeIII}{{\textrm{\bf Type \IIID}}}
\newcommand{\TypeIV}{{\textrm{\bf Type \IVB}}}

\newcommand{\I}{\textrm{I}\xspace}
\newcommand{\II}{\textrm{II}\xspace}
\newcommand{\III}{\textrm{III}\xspace}
\newcommand{\IV}{\textrm{IV}\xspace}

\DeclareFontFamily{OMX}{MnSymbolE}{}
\DeclareSymbolFont{largesymbolsX}{OMX}{MnSymbolE}{m}{n}
\DeclareFontShape{OMX}{MnSymbolE}{m}{n}{
    <-6>  MnSymbolE5
   <6-7>  MnSymbolE6
   <7-8>  MnSymbolE7
   <8-9>  MnSymbolE8
   <9-10> MnSymbolE9
  <10-12> MnSymbolE10
  <12->   MnSymbolE12}{}

\DeclareMathSymbol{\downbrace}    {\mathord}{largesymbolsX}{'251}
\DeclareMathSymbol{\downbraceg}   {\mathord}{largesymbolsX}{'252}
\DeclareMathSymbol{\downbracegg}  {\mathord}{largesymbolsX}{'253}
\DeclareMathSymbol{\downbraceggg} {\mathord}{largesymbolsX}{'254}
\DeclareMathSymbol{\downbracegggg}{\mathord}{largesymbolsX}{'255}
\DeclareMathSymbol{\upbrace}      {\mathord}{largesymbolsX}{'256}
\DeclareMathSymbol{\upbraceg}     {\mathord}{largesymbolsX}{'257}
\DeclareMathSymbol{\upbracegg}    {\mathord}{largesymbolsX}{'260}
\DeclareMathSymbol{\upbraceggg}   {\mathord}{largesymbolsX}{'261}
\DeclareMathSymbol{\upbracegggg}  {\mathord}{largesymbolsX}{'262}
\DeclareMathSymbol{\braceld}      {\mathord}{largesymbolsX}{'263}
\DeclareMathSymbol{\bracelu}      {\mathord}{largesymbolsX}{'264}
\DeclareMathSymbol{\bracerd}      {\mathord}{largesymbolsX}{'265}
\DeclareMathSymbol{\braceru}      {\mathord}{largesymbolsX}{'266}
\DeclareMathSymbol{\bracemd}      {\mathord}{largesymbolsX}{'267}
\DeclareMathSymbol{\bracemu}      {\mathord}{largesymbolsX}{'270}
\DeclareMathSymbol{\bracemid}     {\mathord}{largesymbolsX}{'271}

\makeatletter
\def\horiz@expandable#1#2#3#4#5#6#7#8{%
  \@mathmeasure\z@#7{#8}%
  \@tempdima=\wd\z@
  \@mathmeasure\z@#7{#1}%
  \ifdim\noexpand\wd\z@>\@tempdima
    $\m@th#7#1$%
  \else
    \@mathmeasure\z@#7{#2}%
    \ifdim\noexpand\wd\z@>\@tempdima
      $\m@th#7#2$%
    \else
      \@mathmeasure\z@#7{#3}%
      \ifdim\noexpand\wd\z@>\@tempdima
        $\m@th#7#3$%
      \else
        \@mathmeasure\z@#7{#4}%
        \ifdim\noexpand\wd\z@>\@tempdima
          $\m@th#7#4$%
        \else
          \@mathmeasure\z@#7{#5}%
          \ifdim\noexpand\wd\z@>\@tempdima
            $\m@th#7#5$%
          \else
           #6#7%
          \fi
        \fi
      \fi
    \fi
  \fi}

\def\overbrace@expandable#1#2#3{\vbox{\m@th\ialign{##\crcr
  #1#2{#3}\crcr\noalign{\kern2\p@\nointerlineskip}%
  $\m@th\hfil#2#3\hfil$\crcr}}}
\def\underbrace@expandable#1#2#3{\vtop{\m@th\ialign{##\crcr
  $\m@th\hfil#2#3\hfil$\crcr
  \noalign{\kern2\p@\nointerlineskip}%
  #1#2{#3}\crcr}}}

\def\overbrace@#1#2#3{\vbox{\m@th\ialign{##\crcr
  #1#2\crcr\noalign{\kern2\p@\nointerlineskip}%
  $\m@th\hfil#2#3\hfil$\crcr}}}
\def\underbrace@#1#2#3{\vtop{\m@th\ialign{##\crcr
  $\m@th\hfil#2#3\hfil$\crcr
  \noalign{\kern2\p@\nointerlineskip}%
  #1#2\crcr}}}

\def\bracefill@#1#2#3#4#5{$\m@th#5#1\leaders\hbox{$#4$}\hfill#2\leaders\hbox{$#4$}\hfill#3$}

\def\downbracefill@{\bracefill@\braceld\bracemd\bracerd\bracemid}
\def\upbracefill@{\bracefill@\bracelu\bracemu\braceru\bracemid}

\def\upbrace@expandable{%
  \horiz@expandable
    \upbrace
    \upbraceg
    \upbracegg
    \upbraceggg
    \upbracegggg
    \upbracefill@}
\def\downbrace@expandable{%
  \horiz@expandable
    \downbrace
    \downbraceg
    \downbracegg
    \downbraceggg
    \downbracegggg
    \downbracefill@}

\DeclareRobustCommand{\overbrace}[1]{\mathop{\mathpalette{\overbrace@expandable\downbrace@expandable}{#1}}\limits}
\DeclareRobustCommand{\underbrace}[1]{\mathop{\mathpalette{\underbrace@expandable\upbrace@expandable}{#1}}\limits}
\makeatother

\begin{document}

\begin{titlepage}

\begin{center}
\vspace{1.5cm}

{\Large \bf Recursion Relations for Conformal Blocks}

\vspace{0.8cm}

{\bf Jo\~ao Penedones$^{1,2}$, Emilio Trevisani$^{1,3}$, Masahito Yamazaki$^{4,5}$}

\vspace{.5cm}
{
\small
{\it  $^1$Centro de Fisica do Porto,
 Departamento de Fisica e Astronomia,\\
Faculdade de Ci\^encias da Universidade do Porto,
Porto, Portugal
}\\

\vspace{.3cm}
{\it  $^2$CERN, Theory Group, Geneva, Switzerland }\\

\vspace{.3cm}
{\it  $^3$  ICTP South American Institute for Fundamental Research, IFT-UNESP, S\~ao Paulo, Brazil  
}\\

\vspace{.3cm}
{\it  $^4$Kavli Institute for the Physics and Mathematics of the Universe (WPI), \\
University of Tokyo,  Kashiwa, Japan }\\

\vspace{.3cm}
{\it $^5$School of Natural Sciences, Institute for Advanced Study, Princeton, USA}

}
\end{center}

\vspace{1cm}

\begin{abstract}

In the context of conformal field theories in general space-time dimension, we find all the possible singularities of the conformal blocks as functions of the scaling dimension $\Delta$ of the exchanged operator. In particular, we  argue, using representation theory of parabolic Verma modules, that in odd spacetime dimension the singularities are only simple poles. We discuss how to use this information to write recursion relations that determine the conformal blocks. We first recover the recursion relation introduced in \cite{arXiv:1307.6856} for conformal blocks of external scalar operators. We then generalize this recursion relation for the conformal blocks associated to the four point function of three scalar and one vector operator. Finally we specialize to the case in which the vector operator is a conserved current.

\end{abstract}

\bigskip

\end{titlepage}


\tableofcontents

\section{Introduction}

The conformal bootstrap program \cite{Ferrara:1973yt, Polyakov:1974gs} is a non-perturbative approach to the construction of interacting conformal field theories (CFT), which has made remarkable progress in recent years
since the pioneering work of \cite{arXiv:0807.0004}.
This approach has been successfully applied to 
many conformal field  theories 
\cite{arXiv:0905.2211, arXiv:1009.2087, arXiv:1009.2725, arXiv:1009.5985, arXiv:1109.5176, arXiv:1203.6064, arXiv:1210.4258, arXiv:1304.1803, arXiv:1307.3111, arXiv:1307.6856, arXiv:1309.5089, arXiv:1403.4545, arXiv:1404.0489, arXiv:1406.4814, arXiv:1406.4858, arXiv:1412.4127, arXiv:1412.7541,
arXiv:1502.02033, arXiv:1502.07217, arXiv:1503.02081, arXiv:1504.07997, 
arXiv:1507.05637, arXiv:1508.00012}.

Conformal blocks (CB) are the basic ingredients needed to set up the conformal bootstrap equations. 
They encode the contribution to a four point function from the exchange of a primary operator and all its descendants.
The CBs depend on the spacetime dimension $d$ and on the scaling dimensions and $SO(d)$ irreducible representations labelling the four external  primary operators and the exchanged operator.
Explicit formulas are known only for simple cases like external scalar operators in even spacetime dimension \cite{hep-th/0011040, hep-th/0309180}.
In other cases one has to resort to more indirect methods like, integral representations \cite{Ferrara:1972xe, Ferrara:1972ay, Ferrara:1972uq, Ferrara:1973vz, arXiv:1204.3894, arXiv:1411.7351,  arXiv:1508.02676}, recursion relations that increase the spin of the exchanged operator \cite{hep-th/0011040, arXiv:1108.6194} or efficient series expansions \cite{arXiv:1303.1111, arXiv:1305.1321}.
There is also a general method to increase the spin of the external operators using differential operators \cite{arXiv:1109.6321, arXiv:1505.03750, arXiv:1508.00012}. However, this method cannot change the $SO(d)$ representation of the exchanged operator.
In this paper, we consider a different method  proposed in \cite{arXiv:1307.6856, arXiv:1406.4858}, which generalizes an old idea of Zamolodchikov for Virasoro conformal blocks \cite{Zamolodchikov:1985ie}.
The idea is to consider the CB as a function of the scaling dimension $\D$ of the exchanged operator and analyze its analytic structure in the $\D$ complex plane. As explained in the next section, all poles in $\D$ are associated to the existence of null states in the exchanged conformal family which leads to residues proportional to other CBs. This knowledge leads to recursion relations that can be used to efficiently determine the CBs.

In this paper, we carefully explain the several ingredients that go into the construction of the recursion relations in 3 simple cases.  
Namely, the exchange of a symmetric traceless tensor in the four point function of 4 external scalars in section \ref{sec:scalarCB},  3 scalars and 1 vector in section \ref{sec:onevectorCB} and 3 scalars and 1 conserved current in \ref{sec:onecurrentCB}. 
The case of the vector operator illustrates a new feature that arises when there is more than one CB for a given set of labels of the external and exchanged operators. In this case, the residues in the $\D$ complex plane become linear combinations of  several CBs.

In section \ref{StructureofCFT}, we give a detailed discussion of the structure of general conformal families in 
any spacetime dimension. In particular, we determine all possible singular values of $\D$ for any CB. We argue that these singularities are simple poles in odd spacetime dimension $d$ (and generic non-integer dimension, by analytic continuation in $d$). We conclude with some remarks about the residues of these poles in general spacetime dimension but leave their explicit computation for future work.

\section{Basic Idea}  
We start by summarizing the basic idea that gives rise to  recursion relations for the conformal blocks.
Before that we   introduce some standard CFT notation.

\subsection{CFT Preliminaries}

Let us first begin by setting up our notations and conventions (see \cite{Sofia,Dobrev:1977qv,SpinningCC}). 
Throughout this paper, a symmetric and traceless tensor $\Tcal^{\m_1\ldots \m_l}$ is encoded by a 
polynomial $\Tcal(z)$ by
\beq
 \Tcal(z)\equiv  \Tcal^{\m_1\ldots \m_l} z_{\m_1} \cdots z_{\m_l} \ ,
\eeq
where $z^{\m}$ is a vector in $\mathbb{C}^d$ that satisfy $z \cdot z=0$. We can recover the tensor from the polynomial by the relation
\beq
\Tcal^{\m_1\dots \m_l}  = \frac{1}{l! (h-1)_l} D_z^{\m_1}\cdots D_z^{\m_l}  \Tcal(z)\ .
\label{fTod_appendix}
\eeq
where $ h \equiv d/2$, $(a)_l\equiv \Gamma(a+l)/\Gamma(a)$ is the Pochhammer symbol, and  $D_z^\m$ is the differential operator defined by
\beq
D^\m_z \equiv \left(h-1+z\cdot \frac{\partial}{\partial z} \right) \frac{\partial}{\partial z_\m} -\frac{1}{2} z^\m  
\frac{\partial^2\ }{\partial z \cdot \partial z}\ .
\label{Tod}
\eeq
In this formalism, we denote a primary operator with spin $l$ and conformal dimension $\D$ by
\be 
\Ocal(x,z)=\Ocal^{\mu_1 \dots \mu_l}(x)z_{\m_1}\cdots z_{\m_l} \ .
\ee

The two point function of (canonically normalized) primary operators  is given by
\be \label{def:2ptspin}
\langle \Ocal(x_1,z_1) \Ocal(x_2,z_2) \rangle = \frac{ (z_1 \cdot I(x_{12}) \cdot z_2)^l }{x_{12}^{2\D}} \ ,
\ee
with $x_{12}\equiv x_1-x_2$ and
\be
I_{\m \n}(x) \equiv \eta_{\m \n}-2\frac{x_\m x_\n }{x^2} 
\ .
\ee

The three point function of two scalar primaries $\mathcal{O}_{1}, \mathcal{O}_2$ and one spin $l$ primary operator $\mathcal{O}$ is completely fixed up to an overall constant,
\be
\langle \Ocal_1(x_1)\Ocal_2(x_2) \Ocal(x_3,z_3) \rangle =c_{12\Ocal}\,
\frac{(z_3\cdot y)^l
}{(x_{12}^2)^\frac{\D_1+\D_2-\D+l}{2}
(x_{13}^2)^\frac{\D+\D_{12}-l}{2}
(x_{23}^2)^\frac{\D-\D_{12}-l}{2}}\ ,
\ee
where $\D_{ij}\equiv \D_i-\D_j$, $c_{i j k}$ are the structure constants of the conformal field theory and
$
y^\mu \equiv \frac{x_{13}^\mu}{x_{13}^2} -
\frac{x_{23}^\mu}{x_{23}^2}\ .
$
Notice that this corresponds to the OPE
\be
\Ocal(x,z)\Ocal_1(0) = 
c_{12\Ocal}\, \frac{(-x\cdot z)^l}
{(x^2)^\frac{\D+\D_{12}+l}{2}}\left[\Ocal_2(0) +
{\rm (descendants)} \right]\ .
\label{OPE}
\ee

\begin{figure}[htbp]
\graphicspath{{Fig/}}
\def\svgwidth{7 cm}
\centering
\input{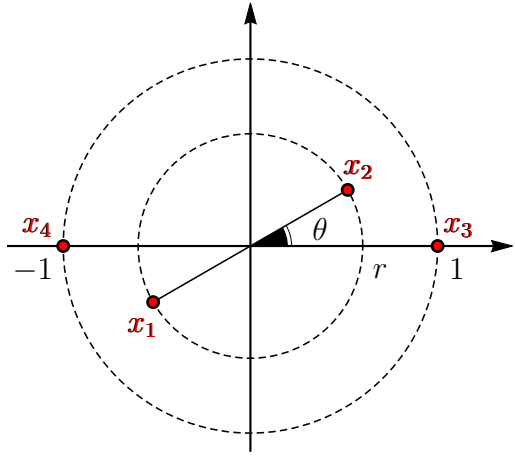} 
\caption{\label{fig:rcoord} Radial coordinates of \cite{arXiv:1303.1111}.}
\end{figure}

Four-point functions depend non-trivially on the conformal invariant cross ratios
\be
u\equiv \frac{x_{12}^2 x_{34}^2}{x_{13}^2 x_{24}^2}\ , \ \ \ \ \ \ \ \ \ \ \ \ \ \ \ \ \ \ 
v\equiv \frac{x_{14}^2 x_{23}^2}{x_{13}^2 x_{24}^2} \ .
\ee
Instead of $u$ and $v$, it will be more convenient in the following to use the radial coordinates $r, \eta$ of \cite{arXiv:1303.1111}.
Using a conformal transformation to map $x_i$ to the configuration shown in figure \ref{fig:rcoord}, we define the radial coordinates by $r$ and $\eta =\cos \theta$.    

To state the exact relation between the radial coordinates $r, \eta$ and the standard cross ratios $u, v$, we first define $z$ and $\bar z$ via $u=z\bar{z}$ and $v=(1-z)(1-\bar{z})$. We can then map $z$ and $\bar z$  to  $r$ and $\eta=\cos \theta$ using
\be \label{radcoord}
re^{i\theta}=\frac{z}{(1+\sqrt{1-z})^2} \ , \qquad re^{-i\theta}=\frac{\bar z}{(1+\sqrt{1-\bar z})^2} \ .
\ee

\subsection{Basic Idea} 
\label{General Idea} 

In this section we summarize the basic idea of this paper. 
To make a cleaner exposition we consider the conformal block of a four point function of scalar primary operators,
\begin{align}
\langle \Ocal_1(x_1) \Ocal_2(x_2)\Ocal_3(x_3) \Ocal_4(x_4) \rangle &=\sum_{\Ocal} c_{12\Ocal} c_{34\Ocal} G_{\D,l}(x_1,x_2,x_3,x_4)  \nonumber \\ 
&=\sum_{\Ocal}%
\raisebox{1.7em}{
$\xymatrix@=6pt{ {\Ocal_1}
\ar@{-}[rd] &  &   & &\Ocal_3 \ar@{-}[ld]   \\  
 & *+[o][F]{} 
\ar@{=}[rr]^{\displaystyle \Ocal }  
& &
*+[o][F]{}  &  \\
\Ocal_2 \ar@{-}[ru]& & & &  \Ocal_4 \ar@{-}[lu]}$
}\ ,    \nonumber 
\end{align} 
where $\Ocal_i$ are scalar primary operators with dimension $\D_i$ and $\Ocal$ is a primary operator with dimension $\D$ and spin $l$ (for clarity of the exposition we first suppress the $l$ indices of $\Ocal$).
Conformal symmetry implies that each $G_{\D,l}(x_1,x_2,x_3,x_4)$ is fixed up to a function of two conformal invariant real variables $r, \eta$ introduced in the previous section: 
\be
G_{\D,l}(x_1,x_2,x_3,x_4)=
\frac{\left(\frac{x_{24}^2}{x_{14}^2} \right)^\frac{\D_{12}}{2}
\left(\frac{x_{14}^2}{x_{13}^2} \right)^\frac{\D_{34}}{2}}{(x_{12}^2)^{\frac{\D_1+\D_2}{2}} 
(x_{34}^2)^\frac{\D_3+\D_4}{2}} \;
g_{\D,l}(r,\eta) \ ,
\label{Gandg}
\ee
where $\D_{ij}\equiv \D_i-\D_j$. It is convenient to express the conformal block  as a sum over states in radial quantization 
\be \label{CBRadialQuantization}
c_{12\Ocal} c_{34\Ocal}G_{\D,l}(x_1,x_2,x_3,x_4) = \sum_{\a \in \Hcal_{\Ocal}} \frac{ \langle 0| \Ocal_1(x_1)\Ocal_2(x_2) | \a \rangle \langle \a| \Ocal_3(x_3)\Ocal_ 4(x_4)|0 \rangle}{\langle \a | \a \rangle} \ ,
\ee
where $\Hcal_{\Ocal}$ is the irreducible representation of the conformal group associated with the primary $\Ocal$ (\emph{i.e.}, it is $\Ocal$ with all its descendants). 

We will study $G_{\D,l}(x_1,x_2,x_3,x_4) $ as a function of $\D$. For some special values $\D=\D^\star_A$,    there exists a descendant state \footnote{Again we suppress the indices of $\Ocal_A$.} 
\be
|   \Ocal_A \rangle \in \Hcal_{\Ocal} \qquad \mbox{ with } 
\left\{
\begin{array}{cl}
\D_A &= \D^\star_A+n_A \\
l_A &
\end{array}
\right. 
\ee 
at the level $n_A$, that becomes a primary. Namely
\be
K_{\nu} |   \Ocal_A \rangle =0
\ee 
for all special conformal generators $K_{\nu}$.
\begin{figure}[t!]
\graphicspath{{Fig/}}
\def\svgwidth{10 cm}
\centering
\input{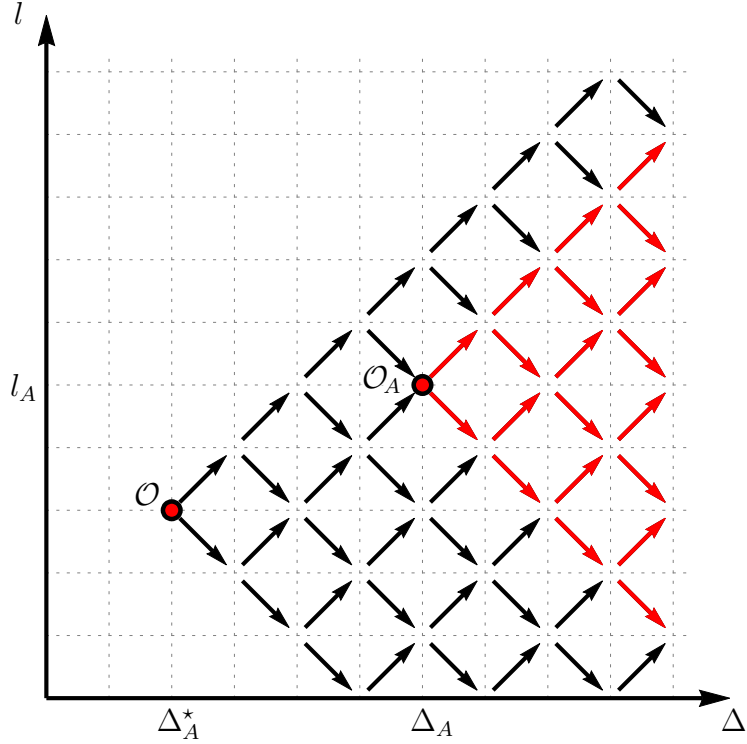} 
\caption{\label{fig:subrep} The conformal representation $\Hcal_\Ocal$ becomes reducible for some special values $\D^\star_A$ of the conformal dimension of $\Ocal$. The descendant state $\Ocal_A$ becomes a primary and the representation $\Hcal_{\Ocal_A}$ only contains null states.
}
\end{figure}
It is easy to show that any state that is both primary and descendant  must have zero norm. Therefore, it follows that the denominator of (\ref{CBRadialQuantization}) becomes zero when $\D=\D^\star_A$,
\be
G_{\D,l} \underset{\D \rightarrow \D^\star_A}{\longrightarrow} \infty \ .
\ee
If  $|   \Ocal_A \rangle$ 
is a null primary state, its descendants are also null and together they form a conformal sub-representation.  It then follows that $G_{\D,l}$ has a pole at $\D = \D^\star_A$ and its residue is proportional to a conformal block, 
\footnote{In some cases, the singularity can be a higher order pole. Moreover, in general the residue of the pole may not be another conformal block. Some of these exceptions are discussed in section \ref{StructureofCFT} but they will not be important for the rest of the paper. }
\be
G_{\D,l}(x_1,x_2,x_3,x_4) \longrightarrow  \frac{R_A}{\D-\D^\star_A}  G_{ \D_A  l_A}(x_1,x_2,x_3,x_4) \qquad \textrm{as} \qquad \D\rightarrow \D^\star_A \ .
\label{eq:GDeltal1}
\ee
The coefficient $R_A$ will be given by three contributions
 \be
 R_A=M_A^{(L)} Q_A M_A^{(R)} \ ,
 \label{QMM}
 \ee
where $Q_A$ comes from the inverse of the norm  $ \langle \alpha | \alpha \rangle$  of the intermediate state $\alpha$, while $M_A^{(L)}$ and $ M_A^{(R)}$ come from the three point functions  $\langle 0| \Ocal_1(x_1)\Ocal_2(x_2) | \a \rangle$ and $\langle \a| \Ocal_3(x_3)\Ocal_4(x_4) | 0 \rangle$. 

To be more explicit we restore the indices and we define the normalization of the primary state $|\Ocal ;z \rangle \equiv  z_{\m_1}\dots z_{\m_l} |\Ocal^{\m_1, \dots, \m_l}\rangle$ with spin $l$ and dimension $\D$, as
\be
\langle \Ocal ;z  |\Ocal ;z\rq{}\rangle = (z\cdot z\rq{})^l \ ,
\ee
and write the norm of the descendant state $|\Ocal_A;z \rangle $ as
\be
\langle \Ocal_A;z |\Ocal_A;z\rq{}\rangle= N_A (z\cdot z\rq{})^{l_A} \ .
\label{norm_OA}
\ee
Since $|   \Ocal_A^{\n_1 \dots \n_{l_{ A}}} \rangle$ is a null state,
$N_A$ has a zero at $\D=\D_A^\star$, so that the inverse of $N_A$ will have a pole at $\D=\D_A^\star$ with some residue $Q_A$.

For what concerns the three point functions, the OPE for a primary operator is given in \eqref{OPE},
while the OPE of a primary descendant operator  takes a similar form, 
except for the   factor $M^{(L)}_A$
\begin{align}
 \Ocal_A(x,z)\Ocal_1(0)&=M^{(L)}_A c_{12 \Ocal }\frac{(-x\cdot z)^{l_A}}
{(x^2)^\frac{\D_A+\D_{12}+l_A}{2}}\left[\Ocal_2(0) +
{\rm descendants} \right]\ . 
\end{align}
In the same way one could define $M^{(R)}_A$ from the OPE involving 
$\Ocal_3$ and $\Ocal_4$.

The idea of the paper is to find  all the poles $\Delta^\star_A$ in $\D$ of the conformal block and the associated primary descendant states $|\Ocal_A^{\n_1 \dots \n_{l_{ A}}} \rangle$. Using this information we will compute the residue $R_A$ of the conformal block at each pole $\Delta^\star_A$.
In addition, we will study the behavior of the conformal block $g_{\D l}$ as $\D \to \infty$.
  To be precise  $g_{\D l}$ has an essential singularity when $\D \rightarrow \infty$.
However, one can define  a regularized conformal block $h_{\D l}(r,\eta)$ such that \cite{arXiv:1307.6856}
\begin{align}
g_{\D l}(r,\eta)&= (4r)^{\D} h_{\D l}(r,\eta) \ ,\label{def:hdeltal}\\
h_{\D l}(r,\eta)&=h_{\infty l}(r,\eta)+\sum_{A} \frac{R_A }{\D-\D^\star_A} (4 r)^{n_A}
h_{\D_A\,l_A}(r,\eta) \ ,
\label{recScalar}
\end{align}
where $h_{\infty l}(r,\eta)=\lim_{\D \rightarrow \infty} h_{\D l}(r,\eta)$ is  finite
and can be computed explicitly (see below).
Equation (\ref{recScalar}) can be used as a recursion relation to determine the conformal block. Notice that if we plug into the right hand side of (\ref{recScalar}) a series expansion for $h_{\D l}(r,\eta)$ correct up to $O(r^k)$,
then the left hand side will be correct up to $O(r^{k+1})$ because $n_A \ge  1$.
Furthermore, for this purpose, we only need to keep  a finite number of terms in the sum over primary descendants $A$.

\section{Conformal Blocks for External Scalar Operators}
\label{sec:scalarCB}
In this section we  rederive  the recursion relation for  conformal blocks of external scalar operators presented in \cite{arXiv:1307.6856,arXiv:1406.4858}. 
As we discussed in section \ref{General Idea},  three ingredients are needed to write such a formula: the knowledge of the data relative to the null primary states (namely $\D^\star_A$, $\D_A$ and $l_A$), the value of the residue $R_A$ and the conformal block at large $\D$. In the following three subsections we will study   each one of these three ingredients and in section \ref{scalarrecrel} we  will put them together 
into a recursion relation.

\subsection{Null States} \label{subsec.nullstates}

In this section we list all the null states
that can appear in the OPE between two scalars. 
A null state is a state $|\psi \rangle$ that is orthogonal to all other states.
Notice that every null state is either a primary descendant, \emph{i.e.} it is a descendant that is annihilated by the generators of special conformal transformations $K_\mu$, or it is a descendant of a primary descendant.
This follows from the fact that if $|\psi\rangle$ is null and $K_\mu|\psi\rangle \neq 0$ then $K_\mu|\psi\rangle  $ is also null and it has lower dimension. We can continue this process until we reach a null state $|\psi_0\rangle$ that is also primary. The original null state $|\psi\rangle$ is then a descendant of  $|\psi_0\rangle$.
Therefore, it is sufficient to look for all primary descendants that arise in a conformal family when the dimension $\Delta$ of the primary varies.
Since any operator that appears in the OPE of two scalar operators has to belong to the traceless and symmetric representation of the rotation group $SO(d)$, it is sufficient to look for primary descendants in this   representation.  

\begin{figure}[t!]
\graphicspath{{Fig/}}
\def\svgwidth{10 cm}
\centering
\input{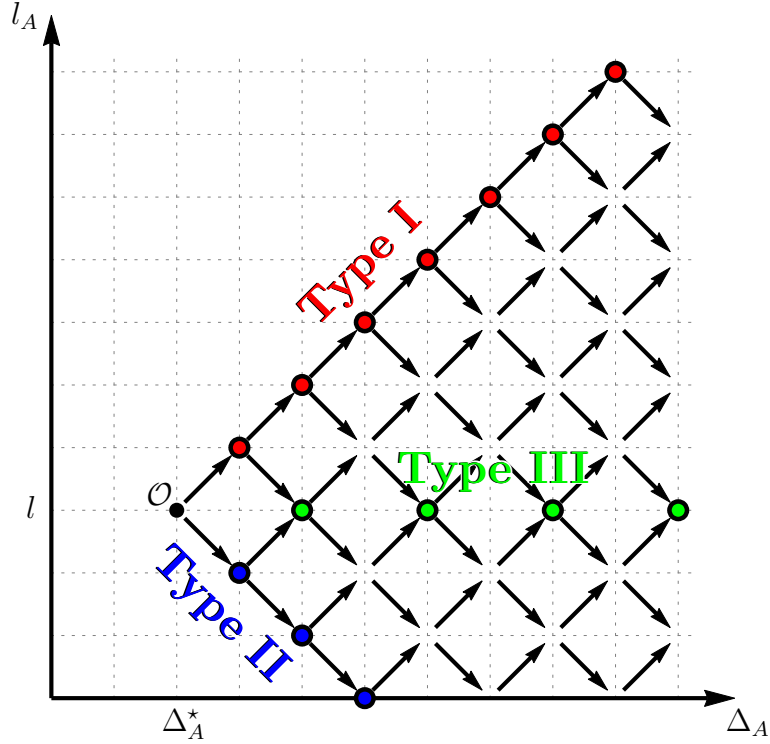} 
\caption{\label{fig:3Types}
The picture shows the traceless and symmetric part of $\Hcal_\Ocal$. Each sequence of arrows creates a descendants of $\Ocal$. When the conformal dimension of $\Ocal$ takes a value $\D^\star_{A}$, one descendant placed at a colored dot becomes a primary state.  There are  three types of primary descendants labeled by an integer $n$ that counts the dots from the left to the right.
}
\end{figure}

Consider traceless and symmetric primary state with spin $l$ 
\be
|\D,l \,;z \rangle \equiv z_{\m_1} \dots z_{\m_l} \Ocal^{\mu_1 \dots \mu_l}(0)|0\rangle\ \equiv \Ocal(z,0)|0\rangle\ .
\ee
normalized as in \eqref{norm_OA}.
We  find (see section \ref{StructureofCFT} for justification)   that, in the OPE of two scalars, all primary descendants 
are of the following 3 types, which we call type \I, type \II and type \III
 (see figure \ref{fig:3Types})\footnote{Type \II and type \III in our paper is type \III and type \II in \cite{arXiv:1307.6856}, respectively.}.  
For each type the null states are labeled by an integer $n$,
which as we will see runs over all positive integers for type \I and type \III, and over a finite set for type \II.
In the following we will collectively denote the type (\I, \II, \III) and the integer $n$ by the label $A$,
\be
A\equiv \textrm{Type},n \ , \qquad 
\left\{
\begin{array}{ccl}
\textrm{Type}&=&\I,\II,\III \\
n &=& 1, 2, \dots 
\end{array}
\right. \ .
\ee
All the primary descendant states can be written as a differential operator $\mathcal{D}_A$ acting on the primary state:
\be
| \D_A,l_A\,; z \rangle = \Dcal_A  |\D,l \,;z \rangle \ .
\label{DA}
\ee

\paragraph{Type \I)}
Type \I states are the
maximal spin descendant at level $n$
\be
|\D+n,n+l\, ;z\rangle=
\Dcal_{\I,n}|\D,l\, ;z \rangle\equiv (z\cdot P)^n |\D,l\, ;z \rangle \ ,
\label{type_I}
\ee
where $P^\mu$ is the generator of translations.
In formula (\ref{norm_I}) in appendix \ref{app.norm} we find that the norm of the  state \eqref{type_I} becomes null when 
\be
\D=\D^\star_{\I, n}\equiv 1-l-n \qquad (n=1,2,\dots) \ .
\ee

\paragraph{Type \II)} 
Type \II states are the minimal spin descendant at level $n \le l$
\be
|\D+n,l-n;z\rangle =\Dcal_{\II,n}|\D,l\, ;z \rangle \equiv
\frac{(D_z \cdot P)^n}{(2-h-l)_n(-l)_n} |\D,l\, ;z \rangle \ ,
\label{type_II}
\ee
where $D_z$ is the  operator introduced in \eqref{Tod} and we recall that $h=d/2$.
In formula (\ref{norm_II}) in appendix \ref{app.norm} we compute its norm and we show that the state becomes null when 
\be
\D=\D^\star_{\II, n}\equiv l+d-1-n \qquad (n=1,2,\dots, l) \ .
\ee

\paragraph{Type \III)} 
Type \III primary descendant states have the same spin as the original primary,
\be
|\D+2n,l\,;z\rangle   =\Dcal_{\III,n}|\D,l\, ;z \rangle \equiv  \Vcal_0 \cdot \Vcal_1 \cdots \Vcal_{n-1}
|\D,l\, ;z \rangle \ , \label{stateIIIother} 
\ee
where 
\be  \label{Vcalj}
\Vcal_j\equiv P^2-2 \frac{ (P\cdot z) (P\cdot D_z)}{ (h+l+j-1) (h+l-j-2)}  \ .
\ee
In appendix  \ref{app.norm} we compute the norm of (\ref{stateIIIother}) in various cases and we conjecture that this descendant state becomes primary when 
\be
\D=\D^\star_{\III, n}=\frac{d}{2}-n  \qquad (n=1,2, \dots) \ .
\ee

In appendix  \ref{app.norm} we computed the norm $N_A$ of the three states and we found that it has a zero at $\D=\D^\star_A$. In what follows, we will only need the residue $Q_A$ of the inverse of the norm at the pole $\D=\D^\star_A$
\be
\frac{1}{N_A} \approx 
\frac{Q_A}{\D- \D^\star_A}\ ,
\qquad \qquad \D \to \D^\star_A\ .
\ee
We found
\begin{align}
&Q_{\I,n}= -\frac{n }{2^n (n!)^2} \ , \\
&Q_{\II,n} =  -\frac{n (-l)_n}{(-2)^n (n!)^2 (2 h+l-n-2)_n} \frac{( h+ l- n-1)}{ (h+l-1)} \ , \\
&Q_{\III,n} =-\frac{n }{(-16)^n (n!)^2 (h-n-1)_{2 n}} \frac{(h+l-n-1)}{ (h+l+n-1)} \ .
\end{align}

\subsection{Residues \texorpdfstring{$R_A$}{RA}}
We shall now compute the OPE coefficient  $M_A$ of the primary descendant  $A$.  
Since for a scalar primary operator $\Ocal(x,z)$ the OPE can be written as (\ref{OPE}),
the OPE of the primary descendant $\Dcal_A \Ocal(x,z)$ is given by
\begin{align}
\Dcal_A \Ocal(x,z)\Ocal_1(0,z_1)&=c_{12\Ocal}\, \Dcal_A \frac{(-x\cdot z)^l}
{(x^2)^\frac{\D+\D_{12}+l}{2}}\left[\Ocal_2(0) +
{\rm descendants} \right]\ ,
\label{DAOO1}
\end{align}
(recall $\mathcal{D}_A$ is the differential operator introduced in \eqref{DA}).
At the same time, since $\Dcal_A \Ocal(x,z)$ is itself a primary operator, 
\eqref{DAOO1}   should also take the form of \eqref{OPE}, up to a possible normalization factor.
We can therefore formulate our problem as follows: we want to find a constant $M_A^{(L)}$ such that
\be
\Dcal_A \frac{(-x\cdot z)^l}
{(x^2)^\frac{\D+\D_{12}+l}{2}} = M_A^{(L)}   \frac{(-x\cdot z)^{ l_A}}
{(x^2)^\frac{\D_A+\D_{12}+ l_A}{2}}
\label{DA-MA}
\ee
(recall that $\D_A$ and  $l_A$ are the conformal dimension and spin of the primary descendant $A$).
Computing (\ref{DA-MA}) for the three types we find (see appendix \ref{App:RAscalarconformalblocks})
\begin{align}
\begin{split}
\label{Mscalar}
M^{(L)}_{\I, n}&= 
(2i)^n \left(\frac{\D+\D_{12}+l}{2}\right)_n \ ,\\
M^{(L)}_{\II, n}&= i^n \frac{(d+l-n-2)_n }{\left(h+l-n-1\right)_n} \left(\frac{\Delta+ \Delta_{12}+2-d-l}{2}  \right)_n \ ,\\
M^{(L)}_{\III, n}&=
(-4)^n \frac{(h-n-1)_l}{(h+n-1)_l}  \left( \frac{\Delta +\Delta_{12}+2-d-l}{2}\right) _n \left( \frac{l+\Delta +\Delta_{12}}{2}\right) _n \ .
\end{split}
\end{align}
With this information we can determine the residue $R_A$  by formula \eqref{QMM} (of course $M^{(L)}_{A}$ should be evaluated at $\D=\D^\star_A$).
This result is consistent with the result of \cite{arXiv:1307.6856,arXiv:1406.4858}%
\footnote{We  use the conventions of \cite{hep-th/0011040, arXiv:1109.6321}. Our convention is related to those of \cite{arXiv:1303.1111} and \cite{arXiv:1307.6856,arXiv:1406.4858} as follows:
\be \nonumber
g_{\D,l}^{\text{\cite{arXiv:1303.1111}}}(u,v)=4^{\D}g_{\D,l}^{\text{\cite{arXiv:1307.6856,arXiv:1406.4858}}}(u,v)=
\frac{(-2)^l \left(\frac{d-2}{2}\right)_l}{(d-2)_l} g_{\D,l}^{\rm here}(u,v)\ .
\ee
}.
\subsection{Conformal Block at Large \texorpdfstring{$\D$}{Delta}}
To compute the large $\Delta$ behavior of the conformal block we need to study the conformal Casimir equation
\be
\mathcal{C} \; G_{\D l}(x_1,x_2,x_3,x_4)= c_{\D l} G_{\D l}(x_1,x_2,x_3,x_4) \ ,\label{casimir_eq}
\ee
where $c_{\D l}\equiv \D(\D-d)+l(l+d-2)$ and $\mathcal{C}$ is the conformal Casimir operator acting on the points $x_1$ and $x_2$, as explained in detail in appendix \ref{App:CasimirEquation}. Equation (\ref{casimir_eq}) is a second order differential equation (since we are using the quadratic Casimir) that can be used to compute the full conformal block $G_{\D l}$ \cite{hep-th/0309180}. The leading term in $\D$ of equation (\ref{casimir_eq}) can be used to find the large $\Delta$ behavior of the conformal block. As it is explained in appendix \ref{App:CasimirEquation}, in this limit  (\ref{casimir_eq})  reduces to a first order differential equation for the conformal block that can be solved up to an integration constant. Moreover we can fix this constant computing the OPE limit ($x_{12}, x_{34}\to 0$) of the conformal block.

To find $h_{\infty l}(r,\eta)$ we rewrite (\ref{casimir_eq}) in the radial coordinates \eqref{radcoord} and use the definitions \eqref{Gandg} and  (\ref{def:hdeltal}) to get an equation for $h_{\D l}(r,\eta)$. If we consider the leading behavior in $\D$ we get a simple differential equation 
\\
\scalebox{0.95}{\parbox{\linewidth}{
\begin{equation*} 
\partial_r h_{\infty l}(r,\eta)= 4 \; \frac{ \eta \, (\Delta _{12} - \Delta _{34})   \left(r^2-1\right)^2
-\eta ^2 r \left((1-2 h) r^2+1\right)
-\frac{r^3+r}{2} \left(h (r^2+1)-2\right)
}{\left(r^2-1\right) \left(r^2-2 \eta  r+1\right) \left(r^2+2 \eta  r+1\right)} \, h_{\infty l}(r,\eta) \ ,
\end{equation*}
}} \\
which can be solved as
\be
h_{\infty l}(r,\eta)=\frac{\left(1-r^2\right)^{1-h}}{\left(r^2-2 \eta  r+1\right)^{\frac{1-\Delta_{12}+\Delta_{34}}{2}} \left(r^2+2 \eta  r+1\right)^{\frac{1+\Delta_{12}-\Delta_{34}}{2}}} f_l(\eta) \ .
\label{h_infty_ansatz}
\ee
Here $f_l(\eta)=\lim_{r\to 0} h_{\infty l}(r,\eta)$ is an integration constant.

In order to determine $f_l(\eta)$, we need to fix the behavior of the conformal block as $r\to 0$, or equivalently
$x_{12}^2, x_{34}^2\to 0$. We thus want to 
compute the leading term of the conformal block of four scalars in the limit $x_{12}^2, x_{34}^2\rightarrow 0$. 
For $x_{12}\rightarrow 0$ one has 
(from \eqref{OPE}) 
\be
\Ocal_1(x_1)\Ocal_2(x_2) \underset{x_{12}\rightarrow 0}{\sim } 
\frac{c_{12\Ocal}}{(h-1)_l l! }\frac{(x_{12}\cdot D_z)^l}
{(x_{12}^2)^\frac{\D_1+\D_{2}-\D+l}{2}} \Ocal(x_2,z) \  .
\label{leadscalarOPE}
\ee 
In the four point function we take the leading OPE (\ref{leadscalarOPE}) both on the left  ($x_{12}\rightarrow 0$) and on the right ($x_{34}\rightarrow 0$). We then use (\ref{def:2ptspin}) to express the remaining two point function,  
\begin{align}
G_{\D,l}(x_i) \underset{x_{12},x_{34}\rightarrow 0}{\sim}\   &\frac{1}{\left(l!\, (h-1)_l \right)^2}
\frac{(x_{12}\cdot D_z)^l   }
{(x_{12}^2)^\frac{\D_1+\D_{2}-\D+l}{2} }
\frac{
(x_{34}\cdot D_{z\rq{}})^l   }
{ (x_{34}^2)^\frac{\D_3+\D_{4}-\D+l}{2}  }
\frac{   (z\cdot I(x_{24}) \cdot z')^l}
{ (x_{24}^2)^{\D}} \ .
\end{align}
We can trivially get rid of the variable $z\rq{}$ 
\be
(x_{34}\cdot D_{z\rq{}})^l   (z\cdot I(x_{24}) \cdot z')^l=  l!\, (h-1)_l (z\cdot I(x_{24}) \cdot x_{34})^l \ ,
\ee
since the  variable $z$ already ensures that the resulting tensor is traceless.
In terms of the radial coordinates we then find the leading contribution for $r \rightarrow 0$ of the conformal block, 
\be \label{rLeadConfBlock}
g_{\D,l}(r,\eta) \underset{r\rightarrow 0}{\longrightarrow}  \frac{l!}{(-2)^l \; (h-1)_l } (4r)^{\D} C_l^{h-1}(\eta) \ ,
\ee
where we used the formula \cite{arXiv:1109.6321}
\be
\frac{(x\cdot D_z)^l (y\cdot z)^l}{l!\, (h-1)_l }=
\frac{l!}{2^l\, (h-1)_l } (x^2 y^2)^{l/2} C_l^{h-1}(\hat x \cdot \hat y) \ , \label{formula:gegenb-projector}
\ee
in which $C_l^{\nu}(x)$ is the Gegenbauer polynomial, and $\hat x\equiv x/(x^2)^{1/2}$. Thus we finally find that
\be
f_l(\eta)= \frac{l!}{(-2)^l\, (h-1)_l }  C_l^{h-1}(\eta) \  .
\ee
\subsection{Recursion Relation} 
\label{scalarrecrel}
We now have all the ingredients that we need to build the scalar conformal blocks using the recursion relation  (\ref{recScalar}). For the sake of clarity  in this section we write down explicitly the recursion relation with all the data that we computed in the previous subsections. The recursion relation is
\begin{align}
g_{\D l}(r,\eta)&= (4r)^{\D} h_{\D l}(r,\eta) \ ,
\\
h_{\D l}(r,\eta)&=h_{\infty l}(r,\eta)+\sum_{A} \frac{R_A }{\D-\D^\star_A} (4 r)^{n_A}
h_{\D_A\,l_A}(r,\eta) \ ,
\label{eq:recscalar}
\end{align}
where the sum over $A$ is a sum over the three types and over $n$ with
\be
\begin{array}{|c | ccc|}
\hline
\phantom{\Big(}		A														&\D^\star_A 					&n_A 										&l_A	 \\ 
\hline
\phantom{\Big(} \mbox{Type \I: }  n=1,2,\dots 	\infty					& 1-l-n 			 				& n 	  	   									 &l + n \\ 
\phantom{\Big(} \quad  \mbox{Type \II: }n=1,2\dots,l \;\; \quad 	&\quad   l+2h-1-n  	\quad & \quad n \quad  						&\quad l - n \quad  \\
\phantom{\Big(} \mbox{Type \III: }n=1,2,\dots \infty \; 				& h-n  							& 2n       			 						 &l       \\ 
\hline
\end{array}
\ee
and $\D_A=\D^\star_A+n_A$.
The residues $R_{A}$ are computed from (\ref{QMM}) and read
\\
\scalebox{0.95}{\parbox{\linewidth}{
\begin{align*}
\left\{
\begin{array}{ll}
\phantom{\Big(}\!\!\!\!R_{\I, n}&= \frac{-n (-2)^n }{  (n!)^2} \left(\frac{\D_{12}+1-n}{2}\right)_n \left(\frac{\D_{34}+1-n}{2}\right)_n \  , \\
\phantom{\Big(}\!\!\!\!R_{\II, n}&= \frac{-n \,l!  }{ (-2)^n (n!)^2 (l-n)!   } \frac{(2h+l-n-2)_n}{\left(h+l-n\right)_n\left(h+l-n-1\right)_n}  	      
\left(\frac{\D_{12}+1-n}{2}\right)_n  \left(\frac{\D_{34}+1-n}{2}\right)_n \ , \\
\phantom{\Big(}\!\!\!\!R_{\III, n}&=\frac{-n (-1)^{n} \left(h-n-1\right)_{2 n}}{(n!)^2 \left(h+l-n-1\right)_{2 n} \left(h+l-n\right)_{2 n}}  \left( \frac{\Delta _{12}-h-l-n+2}{2} \right)_n \left( \frac{\Delta _{12}+h+l-n}{2} \right)_n \left( \frac{\Delta _{34}-h-l-n+2}{2} \right) _n \left( \frac{\Delta _{34}+h+l-n}{2} \right)_n  \ .
\end{array}
\right. 
\end{align*}
}} \\
Moreover the regular part in $\D$ of the conformal block is
\be
h_{\infty l}(r,\eta)=  \frac{\left(1-r^2\right)^{1-h}}{\left(r^2-2 \eta  r+1\right)^{\frac{1-\Delta_{12}+\Delta_{34}}{2}} \left(r^2+2 \eta  r+1\right)^{\frac{1+\Delta_{12}-\Delta_{34}}{2}}} \frac{l!}{(-2)^l\, (h-1)_l } C_l^{h-1}(\eta)\ .
\ee

Although we did not show that we exhausted all possible primary descendants states that could rise to poles of the conformal block, we are confident this is the case for three main reasons. Firstly, because we checked (to several orders in the $r$ series expansion) that \eqref{eq:recscalar}  solves the Casimir equation.
Secondly, because it agrees with the results of \cite{arXiv:1406.4858}.
Thirdly, because the general analysis of conformal families presented in section \ref{StructureofCFT} indicates that we indeed have all the relevant primary descendant states.

\section{Conformal Blocks for One External Vector Operator}
\label{sec:onevectorCB}
We now consider the case of a four point function of three scalars and one vector operator. For tensor fields the three point function is no longer fixed up to a single constant; there are several constants to fix and their number grows with the spin of the operators involved.
This fact can  easily be seen considering the leading OPE of one spin $l$ operator $\Ocal$ with a spin one operator $\Ocal_1$ going into a scalar channel  \cite{Mack:1976pa,OsbornCFTgeneraldim,SpinningCC}
\be \label{OPEwithSPINmain}
\Ocal(x,z)\Ocal_1(0,z_1)\sim \frac{\Ocal_2(0)}{(x^2)^{\a}} \sum_{q=1}^2
c_{12\Ocal}^{(q)} \; t^{(q)}_l( x,z,z_1)  \ ,
\ee
where $\a\equiv \frac{\D+\D_1-\D_2+l+1}{2}$,  $c_{12\Ocal}^{(q)} $ are the OPE coefficients and  $ t^{(q)}_l(x,z,z_1)$   are the two allowed tensor structures (see appendix \ref{CBwithspin})
\begin{align}
\begin{split}
t_l^{(1)}( x,z,z_1)&\equiv (x\cdot z)^{l} \, (z_1 \cdot x) \ ,\\
t_l^{(2)}( x,z,z_1)&\equiv (x\cdot z)^{l-1} \, (z\cdot z_1) \,  x^2\ .
\end{split}
\label{t_five_allowed}
\end{align}
Notice that for $l=0$ only $t_l^{(1)}$ exists.

Let us now consider  the conformal block decomposition of a four point function of three scalar and one vector primary operators $\mathcal{F}_4(\{x_i\},z_1)=
\langle\Ocal_1(x_1,z_1)\Ocal_2(x_2)\Ocal_3(x_3)\Ocal_4(x_4) \rangle$, where 
$\Ocal_1(x_1,z_1)= (z_1)_\m \,\Ocal_1^{\m}(x_1)$.
We can write (see appendix \ref{CBwithspin})
\begin{align}
\label{CBwithSPIN}
\mathcal{F}_4(\{x_i\},z_1)&=\sum_{\Ocal} \sum_{q=1}^2 c_{12\Ocal}^{(q)}  c_{34\Ocal}G_{\D l}^{(q)}(\{x_i\},z_1) \\
&={\sum_{\Ocal} \sum_{q=1}^2} %
\raisebox{1.7em}{$\xymatrix@=6pt{{\Ocal_1}\ar@{-}[rd]& & & &\Ocal_3 \ar@{-}[ld]   \\  
& *+[o][F]{\mbox{\tiny q}}  \ar@{=}[rr]^{\displaystyle \Ocal } & & *+[o][F]{\phantom{\mbox{\tiny q}}}  &  \\
\Ocal_2 \ar@{-}[ru]& & & &\Ocal_4 \ar@{-}[lu]}$} ,  
\end{align}
where the exchanged operator $\Ocal$ belongs to the symmetric traceless representation of $SO(d)$, since all the other representations do not couple to external scalars. 

As in the scalar case we can use conformal symmetry to fix each block $G_{\D l}^{(q)}(\{x_i\},z_1)$ in terms of functions of just two real variables. However in this case the conformal block has one vector index, so it  can be fixed in terms of two scalar functions $g^{(q)}_{\D,l,s}(r, \eta)$ that multiply two independent conformal invariant vector structures (see appendix  \ref{Four Point Function of Vector Operators})
\be 
G_{\D l}^{(q)}(\{x_i\},z_1)=\frac{ 
\left(\frac{x_{24}^2}{x_{14}^2} \right)^{\frac{\D_1-\D_2}{2}}  \left(\frac{x_{14}^2}{x_{13}^2} \right)^{\frac{\D_3-\D_4}{2}}}{(x_{12}^2)^{\frac{\D_1+\D_2}{2}}(x_{34}^2)^{\frac{\D_3+\D_4}{2}}} \sum_{s=1}^{2} g^{(q)}_{\D\, l,s}(r, \eta) Q^{(s)}(\{x_i\},z_1)\ , \label{CB:structuresPhysical}
\ee
where $Q^{(s)}(\{x_i\},z_1)=Q_{\m}^{(s)}( x_1,x_2,x_3,x_4) z_1^{\m}$ is a basis of conformal invariant structures.
A convenient way to generate the structures $Q^{(s)}(\{x_i\},z_1)$ is described in appendix \ref{Four Point Function of Vector Operators} and it is given by 
\be
\label{basisQs1}
\left\{
\begin{array}{l}
{Q}^{(1)}(\{x_i,z_i\})=\textfrak{v}_{1, 23} \ ,\\
{Q}^{(2)}(\{x_i,z_i\})=\textfrak{v}_{1, 34} \ ,
\end{array}
\right.
\ee
 where
\be
\textfrak{v}_{i, jk}\equiv(z_i \cdot \hat x_{i j} )\frac{|x_{i k}|}{|x_{j k}|}-(z_i \cdot \hat  x_{i k} )  \frac{ |x_{i j}|}{|x_{j k}|} \ .
\ee

In the following sections we will explain how to find a recursion relation which determines the conformal blocks $G_{\D l}^{(q)}$. 
\subsection{Null States}
The null states that can propagate are exactly of the same three types that we listed in the scalar case, since they have to belong to traceless symmetric representations. 
\subsection{Residues \texorpdfstring{$R_A$}{RA}\label{subsec:residuesvector}}
In the scalar case we found $R_A$ applying formula \eqref{QMM}. Since $Q_A$ and $M^{(R)}_A$ did not change we now need to compute only $M^{(L)}_A$.

The basic strategy to find $M_A^{(L)}$ is rather similar to the scalar case and it basically boils down to acting with the operators $\Dcal^{A}$  on the OPE as in   \eqref{DAOO1}. 
Concretely  we need to act with $\Dcal^{A}$  on the  OPE  (\ref{OPEwithSPINmain}). Since $\Dcal_A \Ocal(x,z)$ is also a primary operator that belongs to the symmetric and traceless representation, we expect to find as result an OPE of the kind (\ref{OPEwithSPINmain}),
\begin{align} \label{DAOO1SPIN}
\Dcal_A \Ocal(x,z)\Ocal_1(0,z_1)&=\frac{\Ocal_2(0)}{(x^2)^{\a_A}}  \sum_{q\rq{}=1}^2 
  c_{12\Ocal_A}^{(q\rq{})} \; t^{(q\rq{})}_{l_A}(x,z, z_1) 
  \ ,
\end{align}
where $ \a_A\equiv \frac{\D_A +\D_1-\D_2+ l_A+1}{2}$,  $ c_{12 \Ocal_A}^{(q)}$ are new OPE constants and $\D_A, l_A$ are   respectively the conformal dimension and the spin of the primary descendant $\Ocal_A \equiv \mathcal{D}_A \Ocal$. Moreover we can think of (\ref{DAOO1SPIN}) as the action of  $\Dcal^{A}$ on the OPE (\ref{OPEwithSPINmain}), thus we can fix the  OPE constants $c_{12 \Ocal_A}^{(q)}$ in terms of the $c_{12 \Ocal}^{(q)}$ of  (\ref{OPEwithSPINmain}). In the scalar case this operation was simple since there was just one OPE coefficient, while in this case the computation is slightly more involved since the two coefficients can mix. In fact the action of $\Dcal^{A}$ on each structure of the OPE (\ref{OPEwithSPINmain}) can generate the other one, therefore we expect to find a $2 \times 2$ matrix   $M_{A}^{(L)}$ such that
\be
\qquad \qquad c_{12 \Ocal_A}^{(q\rq{})}  =\sum_{q=1}^2  c_{12 \Ocal}^{(q)} \left(M_{A}^{(L)}\right)_{ q q\rq{}} \ .
\ee
In practice we can find $M_A^{(L)}$ performing the following computation
\be \label{DAt=MAt}
\Dcal_{A}\frac{t^{(q)}_l(x,z, z_1)}{(x^2)^{\a}} =\sum_{q\rq{}=1}^2 \left(M_{A}^{(L)}\right)_{q q\rq{}} \frac{t_{l_A}^{(q\rq{})}(x,z, z_1)}{(x^2)^{\a_A}}  
\ .
\ee
 As an example, here we show the result for type \I, 
\be
M^{(L)}_{\I,n}=(-2i )^n (\alpha )_{n-1}\left(
\begin{array}{cc}
 n+\alpha -1 & -\frac{n}{2} \\
 0 & \alpha -1 \\
\end{array}
\right) \ .
\ee
The result for the other types is presented in formula (\ref{MAfinalresult}) in appendix \ref{App:The residue RA for vector conformal blocks}). 

\subsection{Conformal Block at Large \texorpdfstring{$\D$}{Delta}}
\label{ConformalBlockAtLargeDelta}

The goal of this section is to find the leading behavior in $\D$ of $g^{(q)}_{\D,l,s}(r, \eta)$. To achieve this goal we want to apply the same procedure we used in the scalar case. In particular we want to solve the leading term in $\D$ of the Casimir equation (see appendix (\ref{App:CasimirEquation}))
\be
\mathcal{C} \; G_{\D l}^{(q)}(\{x_i\},z_1)= c_{\D l} G_{\D l}^{(q)}(\{x_i\},z_1) \label{casimir eq spin}
\ee
with the appropriate leading behavior of $G_{\D l}^{(q)}$  when $x_{12}, x_{34} \rightarrow 0$. 

We consider the Casimir equation at the leading order in $\D$ replacing  the form (\ref{CB:structuresPhysical}) and $g^{(q)}_{\D,l,s}(r, \eta)=(4r)^{\D} h^{(q)}_{\D,l,s}(r, \eta)$ in equation (\ref{casimir eq spin}). As explained in appendix (\ref{App:CasimirEquation}) we obtain two coupled first order differential equations in the variable $r$.
 We consider the following ansatz for the leading term in $\D$ of $h^{(q)}_{\D,l,s}(r, \eta)$:
\begin{align} \label{hss}
&h^{(q)}_{\infty ,l,s}(r,\eta)=\Acal^{\D_{12},\D_{34}}(r,\eta) \sum_{t=1}^2 F_{s}^{\phantom{s}t}(r,\eta) f^{(q)}_{l,t}(\eta)  \ , \\
&\Acal^{\D_{12},\D_{34}}(r,\eta)= \frac{ \left(1-r^2\right)^{-h}}{ \left(1+r^2-2 r \eta \right)^{\frac{\D_{34}-\D_{12}}{2}} \left(1+r^2+2 r \eta \right)^{1+\frac{\D_{12}-\D_{34}}{2}} }  \ ,
\label{A_ansatz}
\end{align}
where $F_{s}^{\phantom{s}t}(0,\eta)=\d_{s,t}$ so that $h^{(q)}_{\infty,l,s}(0,\eta)=f^{(q)}_{l,s}(\eta)$. The overall function $\Acal^{\D_{12},\D_{34}}(r,\eta)$ depends on the external data, and the expression in \eqref{A_ansatz} is inspired by its scalar counterpart \eqref{h_infty_ansatz}, and is chosen so that the differential equation simplifies. The Casimir equation of course will fix only the matrix $F_{s}^{\phantom{s}t}$, while to fix $f^{(q)}_{l,s}(\eta)$ we need to specify the initial conditions as $r\to 0$. Using the ansatz (\ref{hss}), the two coupled  differential equations take the form
\begin{align}
 \label{2CoupledPDEs}
\partial_r F_s ^{\phantom{s}t}(r,\eta)&= \sum_{s\rq{}=1}^{2} \mathcal{M}_{s\, s\rq{}}(r,\eta) F_{s\rq{}}^{\phantom{s}t}(r,\eta) \ , \\
\mathcal{M}(r,\eta)&=\frac{2}{\left(r^2-1\right) \left(r^2+2 \eta  r+1\right)}\left(
\begin{array}{cc}
 r \left(r^2+2 \eta  r-1\right) & 2 r \eta  \\
 r^2+1 & r^3+\eta  r^2+r-\eta  \\
\end{array}
\right) \ .
\end{align}
The label $t=1,2$ of $F_s ^{\phantom{s}t}$ parametrize the two  independent solutions of the two coupled linear differential equations in \eqref{2CoupledPDEs}. The result is
\be
F(r,\eta) =
\left(
\begin{array}{cc}
 r^2+1 & -2 r^2 \eta  \\
 -2 r & -r^2+2 \eta  r+1 \\
\end{array}
\right) \ .
\ee

To find the functions $f^{(p)}_{l, s}(\eta) = h^{(p)}_{\infty,l,s}(0,\eta)$ we need to study the leading behavior of $G_{\D l}^{(p,q)}$ for $r \rightarrow 0$.  This limit corresponds to the leading   contribution   in the left and the right OPE in the four point function. In particular one can completely fix the two functions $f^{(q)}_{l,s}(\eta)$ using 
\begin{align}
\!\!\! \frac{ t^{(q)}_l( \hat x_{12}, 
   I(x_{24})\cdot D_z ,I(x_{12}) \cdot  z_1) (- \hat{x}_{34} \cdot z)^l
}{l! (h-1)_l}
 \approx  \sum_{s} f^{(q)}_{l,s}(\eta) Q^{(s)}(\{x_i,z_i\}) \ ,\label{smallr:eq.forf}
\end{align}
where the left hand side arises from  the leading OPE limit of the four point function and the right hand side comes from the  $r \rightarrow 0$ limit of (\ref{CB:structuresPhysical}). More details about this formula  can be found in appendix \ref{Conformal_Block_at_large_Delta}.
The result is
\be
\!\! \begin{array}{lll}
&f^{(1)}_{l,1}(\eta)=-\dfrac{ l! C_l^{(h-1)}(\eta )}{2^{l}(h-1)_l} \ , &f^{(1)}_{l,2}(\eta)=0 \ , \\
&f^{(2)}_{l,1}(\eta)=-\dfrac{ (l-1)! \big(2 \eta  (h-1) C_{l-1}^{(h)}(\eta )+l C_l^{(h-1)}(\eta )\big)}{ 2^{l}  (h-1)_l} \ , \;
 &f^{(2)}_{l,2}(\eta)=- \dfrac{(l-1)! C_{l-1}^{(h)}(\eta )}{2^{l-1} (h)_{l-1}} \ .
\end{array}
\ee
\subsection{Recursion Relation} 
We now have  all the ingredients to write the recursion relation for all the conformal blocks of one vector operators and three scalars in any dimensions:
\begin{align}
g^{(q)}_{\D l,s }(r,\eta)&=(4 r)^\D h^{(q)}_{\D l,s}(r,\eta) \ , \nonumber  \\
h^{(q)}_{\D l,s}(r,\eta)&=h^{(q)}_{\infty l,s}(r,\eta)+\sum_{q\rq{}=1}^2\sum_{A} \frac{(R_A)_{q q\rq{}} \; (4 \, r)^{n_A}}{\D-\D^\star_A} h^{(q\rq{})}_{\D_A l_A,s}(r,\eta) \ ,
 \label{RecRelSpin+}
\end{align}
where $\D$ and $l$ are respectively the conformal dimension and spin of the exchanged primary and
\be
(R_A)_{q q\rq{}}= (M_A^{(L)})_{q q\rq{}} Q_A  M_A^{(R)} \ .
\ee

\section{Conformal Blocks for One External Conserved Current}
\label{sec:onecurrentCB}
Let us next consider the case where the external vector operator is a conserved current.
The OPE between a conserved current and a scalar is specified by a single tensor structure
\be \label{OPECEeven}
\Ocal(x,z)\Ocal_1(0,z_1)\sim \frac{\Ocal_2(0)}{(x^2)^{\a}} c_{12 \Ocal} \t_{\D l}(x,z,z_1) \ ,
\ee
where $\a=\frac{\D -\D_2+ l+d}{2}$ because the operator $\Ocal_1$ is a conserved current with $\D_1=d-1$. In fact, imposing conservation $\partial_{z_1}\cdot\partial_{x_1} \Ocal_1(x_1,z_1)=0$ on the vector OPE (\ref{OPEwithSPINmain}), we find that $\tau_{\D l}$ is a specific linear combination of the two structures in (\ref{t_five_allowed}),
\begin{align} 
\t_{\D l} &=\sum_{q=1}^2  t_{l }^{(q)} (\w_{\D l})_{ q }  \qquad(l>0) \ , \\
 \w_{\a l}&=(2 (\alpha -1) ,-2 \alpha +2 h+l) \ .  \label{matricesw}
\end{align}
When $l=0$ the second structure does not exist and when we impose conservation we find that there is no allowed $\t_{\D 0}$ unless $\D=\D_2$. We will see that the recursion relation   for $l>0$  does not couple to the case $l=0$ and $\D=\D_2$ so we can compute it separately.

The conformal block for conserved current will be then defined as 
\be 
\tilde G_{\D l}(\{x_i\},z_1)=\frac{ 
\left(\frac{x_{24}^2}{x_{14}^2} \right)^{\frac{\D_1-\D_2}{2}}  \left(\frac{x_{14}^2}{x_{13}^2} \right)^{\frac{\D_3-\D_4}{2}}}{(x_{12}^2)^{\frac{\D_1+\D_2}{2}}(x_{34}^2)^{\frac{\D_3+\D_4}{2}}} \sum_{s=1}^{2} \tilde g_{\D  l,s}(r, \eta) Q^{(s)}(\{x_i\},z_1)\ , 
\ee
where 
\be
\tilde G_{\D l} \equiv \sum_{q=1}^2 (\w_{\D l})_{ q } G_{\D l}^{(q)} \ , \qquad \tilde g_{\D l,s} \equiv \sum_{q=1}^2 (\w_{\D l})_{ q } g_{\D l,s}^{(q)} \ .
\ee
Therefore, the conformal blocks for a conserved current are given by a linear combination of the conformal blocks for a vector operator.
Nevertheless in the next section,  we explain how to obtain a   recursion relation, which directly gives  
\be 
\label{def:htilde}
\tilde h_{\D l,s}(r,\eta)= \sum_{q=1}^2 (\w_{\D l})_{q} h^{(q)}_{\D, l,s}(r,\eta) \ .
\ee

\subsection{Residues \texorpdfstring{$R_A$}{RA}}
The action of $\Dcal_A$ on the OPE (\ref{OPECEeven}) gives back structures of the form (\ref{OPECEeven}), namely
\begin{align}
\Dcal_A \Ocal(x,z)\Ocal_1(0,z_1)&=   \frac{\Ocal_2(0,\partial_{z_2})}{(x^2)^{\a_A}} \;  c_{12 \Ocal} \; \m^{L}_{A} \t_{\D_A l_A}(x,z, z_1,z_2)  \ ,
\end{align}
where $\a_A\equiv \frac{\D  -\D_2+ l_A+ d}{2}$. Now, since there exists just one allowed structure, $ \m^{L}_{A}$ is just a scalar function. Moreover it can be easily computed knowing $M_{A}$:
\be \label{define_mA}
\m_A \left(\w_{\D_A l_A} \right)_q = 
\sum_{q'=1}^2 \left(\w_{\D l}\right)_{q'}  
\left(M_A\right)_{q'q}   \ ,\qquad
\qquad q=1,2 \ ,
\ee
where $\w_{\D l}$ are defined in (\ref{matricesw}). Notice that the two equations in  (\ref{define_mA}) define the same value $\m_A$. This is a non trivial consistency condition for the matrices $M_A$ given in section \ref{subsec:residuesvector}.
We obtain for the three types
\begin{align*}
&\m^{L}_{\I, n}=(- 2 i)^n \left(\frac{\Delta_{12}-n}{2}\right)_n\ , \\
&\m^{L}_{\II, n}  =(-i)^n\frac{ (l-n) }{l }\frac{ (-2 h-l+2)_n}{(-h-l+2)_n}\left(\frac{\Delta_{12}-n}{2}\right)_n\ , \\
&\m^{L}_{\III, n}  =4^n \frac{ (h-n)_{2 n} }{(h+l-n-1)_{2 n}} \left(\frac{h+l-n+\Delta_{12}-1}{2}\right)_n \left(\frac{h+l-n-\Delta_{12}+1}{2} \right)_n\ ,
\end{align*}
where $\D_{12}\equiv d-1-\D_2$.

\subsection{Conformal Block at Large \texorpdfstring{$\D$}{Delta}}

We next want to find the large $\D$ behaviour of   $\tilde h_{\D,l,s}$ defined in (\ref{def:htilde}). 
The function  $\tilde h_{\D,l,s}$   is a linear combination of the functions $h^{(q)}_{\D,l,s}$ that are regular at $\D=\infty$ but the coefficients $ (\w_{\D l})_{q}$ grow linearly in $\D$. This means that we need to compute the sub leading behaviour of $h^{(q)}_{\D,l,s}$ at large $\D$ in order to determine the terms of order $\D$ and $\D^0$ in $\tilde h_{\D,l,s}$.
 To do so we  change the ansatz (\ref{hss}) for the conformal block at large $\D$ to  
\be \label{nexttoleadingansatzh}
h^{(q)}_{\D, l,s}(r,\eta)=  \Acal^{\D_{12},\D_{34}}(r,\eta) \sum_{t=1}^2 F_{s}^{\phantom{s}t}(r,\eta) \left[ f^{(q)}_{l,t}(\eta)+\frac{1}{\D}  \hat f^{(q)}_{l,t}(r,\eta)+ O\left( \frac{1}{\D^2}\right) \right] \ ,
\ee
where $\Acal$, $F$ and $f$ are defined in section \ref{ConformalBlockAtLargeDelta} so that they solve   the Casimir equation at  leading order in the large $\D$ expansion. 
Plugging the ansatz (\ref{nexttoleadingansatzh}) in the Casimir equation,
we find an equation 
of the form $\partial_r \hat f^{(q)}_{l,t}(r,\eta) $ equal to an explicit function of $r$. This can be trivially integrated and the integration constant can be fixed considering that $\hat f^{(q)}_{l,t}(0,\eta)=0$ for consistency with
the leading OPE. 
The final result $\tilde h^{\infty}_{\D,l,s}(r,\eta)$ is then achieved keeping the terms of order $\D$ and $\D^0$ in $ \sum_{q=1}^2 (\w_{\D l})_{q} h^{(q)}_{\D, l,s}(r,\eta) $. The result is
written in formulas (\ref{htildeinfinity1}) and  (\ref{htildeinfinity2}) in appendix \ref{App.ConsCurr}.

\subsection{Recursion Relation} 
Collecting the results above, we can give a simplified version of (\ref{def:htilde}) that computes directly the block of a conserved current without passing through the vector case:
\begin{align}
\tilde g_{\D l,s}(r,\eta)&=(4 r)^\D \tilde h_{\D l,s}(r,\eta) \ ,\\
\tilde h_{\D l,s}(r,\eta)&=\tilde h^{\infty}_{\D,l,s}(r,\eta)+ \sum_{A} \frac{(\r_A) \; (4 \, r)^{n_A}}{\D-\D^\star_A} \tilde h_{\D_A,l_A,s}(r,\eta)  \qquad (l>0) \ ,\label{rechtilde}
\end{align}
where
\be
\r_A=\m^L_A Q_A M^R_A \  
\ee
and $\tilde h^{\infty}_{\D,l,s}(r,\eta)$ contains terms linear and constant in $\Delta$ and is given explicitly in appendix \ref{App.ConsCurr}.
Notice that when we consider $l>0$, the formula never couples to any conformal block with $l=0$. In fact the only way to do so would be through a descendant of type \II at the level $l$, but $\m^L_{\II, l}=0$.  One can check that (\ref{rechtilde}) agrees with (\ref{def:htilde}) at the first orders in the $r$ expansion. For $l=0$ it does not exist the conformal block $\tilde g_{\D l,s}$ unless $\D=\D_2$. To study  $\tilde g_{\D_2 0,s}$ one can use formula (\ref{def:htilde}). 

The blocks computed in (\ref{rechtilde}) are not completely independent, and in fact they have to satisfy the conservation condition
\be
\partial_{x_1} \cdot \partial_{z_1} \tilde{G}_{\D l}(\{x_i\},z_1) = 0 \ .
\ee
This gives a constraint for the functions $\tilde h_{\D l,1}(r,\eta)$ and $\tilde h_{\D 
l,2}(r,\eta)$ that we can schematically write as
\be
\textfrak{D}_{\D}\left[ \tilde h_{\D l,s}(r,\eta)\right]=0 \ , \label{ConservationConditionCB}
\ee
where $\textfrak{D}_{\D}$ is explicitly defined in equation (\ref{frakD}) in appendix \ref{App.ConsCurr}. 
Using the identity
\be
\textfrak{D}_{\D}\left[
r^{n_A} \tilde h_{ \D_A l_A,s}(r,\eta)\right]
=
r^{n_A}\textfrak{D}_{\D+n_A}\left[\tilde h_{\D_A l_A,s}(r,\eta)\right] 
\xrightarrow[\D \to \D_A^\star]{}
0  \ ,
\ee
we conclude that applying $\textfrak{D}_{\D}$ to (\ref{rechtilde}) removes all the poles in $\Delta$. One can also check that 
\be
\textfrak{D}_{\D}\left[\tilde h^{\infty}_{\D l,s}(r,\eta)\right] =O(\D^0) \ .
\ee
This shows that the recursion relation (\ref{rechtilde}) respects conservation up to a term independent of $\Delta$. 
This general strategy is not sufficient to show that this term is actually zero but we have  checked this up to $O(r^7)$ in the series expansion in $r$.

\section{Structure of Conformal Families}
\label{StructureofCFT}

In the previous sections, we saw that poles of the conformal blocks arise when the dimension of the exchanged operator is chosen such that there are null states in its conformal family.
In this section we discuss the general structure of conformal families and a precise mathematical  criterion for the absence of null states.
In order to state this  criterion we must first write the conformal algebra in the Cartan-Weyl basis.
 
\subsection{Conformal Algebra in Cartan-Weyl Basis}
The conformal group of a $d$ dimensional Euclidian CFT is isomorphic to  $SO(d+1,1)$. 
The algebra is generated by $D, P_\m, K_\m, J_{\m \n}$, which generate respectively dilatation, translations, special conformal transformations and rotations in $SO(d)$. The  commutation relations are
\be \label{ConformalAlgebra}
\begin{array}{lllllll}
&[D,P_\mu] &=& i P_\mu \ , \qquad\qquad\qquad\qquad\qquad \ \, &[D,K_\mu]&=&-iK_\mu \ , \\ 
&[P_\mu,J_{\nu\rho}] &=&  i(\eta_{\mu\nu}P_\rho-\eta_{\mu\rho}P_\nu) \ , 
 \qquad \qquad &[K_\mu,J_{\nu\rho}]&=&i(\eta_{\mu\nu}K_\rho-\eta_{\mu\rho}K_\nu) \ ,   \\
& [K_\mu,P_\nu ]& =& 2i (\eta_{\mu \nu} D- J_{\mu\nu}) \ , &&& 
\\
& [J_{\m \n},J_{\r \s}]&=& i(\eta_{\n \r} J_{\m \s} \pm \textrm{perm}) \ . &&& 
\end{array}
\ee
A bosonic primary state can be written  as  $|\D, \nu_{1} \dots \nu_{l} \rangle$, where $\D$ is the conformal dimension and the $l$ indices correspond to an irreducible tensor of $SO(d)$ (possibly with mixed symmetry properties). 
In this tensor representation, the action of the generators is given by
\begin{align}
&D |\D, \nu_{1} \dots \nu_{l} \rangle = i \D |\D, \nu_{1} \dots \nu_{l} \rangle\ ,\qquad
K_\mu |\D, \nu_{1} \dots \nu_{l} \rangle =0\ ,
\label{eq:actionDK}\\
&J_{\a \b} | \D,\nu_{1} \dots \nu_{l} \rangle =
\sum_{k=1}^l [M_{\a \b}]^{\nu_k}_{\ \mu}\ 
| \D,\nu_{1} \dots \nu_{k-1} \, \mu \, \nu_{k+1} \dots \nu_{l}\rangle\ ,
\label{eq:actionJ}
\end{align}
where
\be
[M_{\a \b}]^{\nu}_{\ \mu} = i \left(
\d^{\nu}_\b \eta_{\a \mu} -\d^{\nu}_\a \eta_{\b \mu}
\right)\ .
\ee
The action of $P_\mu$ creates descendants.

Consider first the case of odd dimension $d=2N+1$ and introduce $d$ auxiliary vectors 
\be
z_j^{\eta}\equiv\frac{1}{\sqrt{2}}(\underbrace{0,0}_{1},\dots ,\underbrace{0,0}_{j-1}, \underbrace{1, - \eta i}_{j} , \underbrace{0,0}_{j+1},\dots,\underbrace{0,0}_{N},0) \ ,  \qquad  \qquad z_{N+1}\equiv (0,\dots,0,1) \ ,
\ee
with $\eta=\pm$ and $j=1,2,\dots,N$.
We can then formulate the conformal algebra in a Cartan-Weyl basis:
\begin{align}
&
\left\{ 
\begin{array}{c l}
E_{\a_{0}^{+}}&\equiv K \cdot z_{N+1} \ ,  \\
E_{\a_{0}^{-}}&\equiv P \cdot z_{N+1}  \ ,   \\
E_{\a_{j}^{\eta}}&\equiv \sqrt{2} \; z_j^{\eta} \cdot J \cdot z_{N+1}   \ ,   
\end{array}
\right.  \qquad
\left\{
\begin{array}{c l}
E_{\a_{0 j}^{+ \eta}}&\equiv \frac{1}{\sqrt{2}} \; K \cdot z_j^{\eta} \ ,   \\
E_{\a_{0 j}^{- \eta}}&\equiv \frac{1}{\sqrt{2}} \; P \cdot z_j^{\eta}  \ , \\
E_{\a_{jk}^{\eta_1\eta_2}}&\equiv z_j^{\eta_1} \cdot J \cdot z_k^{\eta_2}  \ ,  
\end{array}
\right. \qquad 
\left\{
\begin{array}{c l}
H_0&\equiv i D  \ ,  \\
H_j&\equiv  i \; z_j^- \cdot J \cdot z_j^+ \ ,  \\
\end{array}
\right.
\label{eq:CartanWeylbasis}
\end{align}
with $1\le j<k\le N$.
The generators $H_{j=0,1,\dots,N}$ commute among themselves and form the Cartan subalgebra.
From this definition and using the commutation relations (\ref{ConformalAlgebra}) it is easy to check  that 
\be \label{CartanWeylAlgebra}
[H_k,E_\a]=(\a)^k E_\a\ , \qquad
[E_\a,E_{-\a}]=\frac{2}{\langle \a,\a \rangle}(\a)^k H_k\ .
\ee
where $\a$ stands for any root of $\so(d+2)$ and 
$(\a)^k$ denotes its $k$-th coordinate.
The root system of $\so(d+2)$ is given by 
$(N+1)$-dimensional vectors
$\a_{j}^{\eta}$ and $\a_{ij}^{\eta_1\eta_2}$,
\be
 \a_{j}^{\eta}=(\underbrace{0}_{0},\dots, 0,\underbrace{\eta}_{j},0,\dots,\underbrace{0}_{N}) \ , \qquad \a_{jk}^{\eta_1 \eta_2 }=(\underbrace{0}_{0},\dots, 0, \underbrace{\eta_1}_{j},0,\dots,0, \underbrace{\eta_2}_{k},0,\dots, \underbrace{0}_{N}) \ . \label{RootsSO2N+1}
\ee
 
In the tensor representation, a generic primary state is defined in terms of $\D$ and a Young tableaux. 
This is specified by a set of integers $(l_1,\dots, l_N)$, where $l_k$ represents the number of boxes of the $k$-th row. Every box is filled with one tensor index and the indices in the same row are symmetric   while the ones in the columns are antisymmetric. We also remove all possible traces of the tensor.
Contracting all the indices of the $i-$th row with vectors $z_i^+$, we obtain a highest weight state
 $| \l  \rangle  $, where $\l\equiv (-\D,l_1,\dots, l_N)$, 
\ytableausetup{centertableaux,boxsize=1.8 em}
\be
{
 \begin{ytableau}
\m^{1}_1 & \cdots&\cdots&\cdots& \cdots&\cdots&\; \m^{1}_{l_1} \\
\m^{2}_1 & \cdots&\cdots&\cdots& \; \m^{2}_{l_2} \\
\none[\vdots] & \none[\vdots]&\none[\vdots] \\
\m^{N}_1 & \cdots& \; \m^{N}_{l_{ N}} 
\end{ytableau} 
} \qquad \Longrightarrow \qquad
{
 \begin{ytableau}
z^+_1 & \cdots&\cdots&\cdots& \cdots&\cdots&\; z^+_1  \\
z^+_2  & \cdots&\cdots&\cdots& \; z^+_2  \\
\none[\vdots] & \none[\vdots]&\none[\vdots] \\
z^+_N  & \cdots& \; z^+_N  
\end{ytableau} 
}
  \; .
\ee
Using   (\ref{eq:CartanWeylbasis}) and (\ref{eq:actionDK}-\ref{eq:actionJ}), it is easy to show that $|\l\rangle$ is annihilated by all positive roots
\begin{align}
&E_{\a_{j}^{+}} |  \l \rangle=0 \ ,\qquad \qquad E_{\a_{ij}^{+ \eta}} |  \l   \rangle =0  \ .
\end{align}
Moreover, one can read off the weights by acting with $H_i$,
\begin{align} \label{CartanEigenvalues}
H_0 |\l \rangle&=-\D|\l \rangle \ , \\
H_i |\l \rangle&=l_i |\l \rangle \qquad (1\le i\le N) \ .
\end{align} 

The quadratic Casimir is given by
\be
\mathcal{C}=\sum_{k=0}^N H_k H_k +
 \frac{1}{2}\sum_{\a \in \Phi}  \langle \a,\a \rangle \, E_\a E_{-\a} \ ,
\ee
where $\Phi$ stands for the set of all roots of $\so(d+2)$. This can be easily evaluated on the highest weight state $|\l \rangle$,
\be
\mathcal{C} |\l \rangle = (\l,\l+2\rho) |\l \rangle =
\left[
\D(\D-d)+\sum_{i=1}^N l_i(l_i+d-2i)
\right]|\l \rangle \ ,
\ee
where the Weyl vector $\rho$
\be
\rho\equiv \frac{1}{2}\sum_{\a \in \Phi^+} \a
\ee
is the half sum of the positive roots.
In this case, it is given by
\be
 \r =\left(N+\frac{1}{2},N-\frac{1}{2},\dots,
 \frac{3}{2},\frac{1}{2}\right) \ .
\ee

The analysis is similar for even dimensions ($d=2N$).  The Cartan subalgebra is the same but we do not have the vector $z_{N+1}$ and the associated generators $E_{\a_{k}^{\eta}}$. This difference has important consequences when we define the highest weight states. Since now $E_{\a_{j}^{+}}$ is absent, we can also build an highest weight state contracting the indices of the last line of the Young tableau with $z_N^-$. 
As a result, we obtain two  highest weight representations associated to each Young tableau with $l_N>0$,
\ytableausetup{centertableaux,boxsize=1.8 em}
\be
{
 \begin{ytableau}
z^+_1 & \cdots&\cdots&\cdots& \cdots&\cdots&\; z^+_1  \\
z^+_2  & \cdots&\cdots&\cdots& \; z^+_2  \\
\none[\vdots] & \none[\vdots]&\none[\vdots] \\
z^+_N  & \cdots& \; z^+_N  
\end{ytableau} 
} \ ,
 \qquad  \qquad 
{
 \begin{ytableau}
z^+_1 & \cdots&\cdots&\cdots& \cdots&\cdots&\; z^+_1  \\
z^+_2  & \cdots&\cdots&\cdots& \; z^+_2  \\
\none[\vdots] & \none[\vdots]&\none[\vdots] \\
z^-_N  & \cdots& \; z^-_N  
\end{ytableau} 
} \ .
\ee
It is easy to see that the eigenvalue of $H_N$ is given by $l_N$ for the first case and by $-l_N$ in the second case.
Therefore, we will label these two representations with the weight vectors $\l=(-\D,l_1,\dots,l_N)$ and
$\l=(-\D,l_1,\dots,-l_N)$.

\subsection{Parabolic Verma Modules}
\label{Section:PVM}

The mathematical concept of parabolic Verma module applies precisely to the conformal families relevant for CFT. Understanding this connection
and using   a theorem of Jantzen \cite{Jantzen}, we will be able to find the general conditions for the absence of primary descendants. We shall follow the notation of the book \cite{Humphreys}.

To define a parabolic Verma module, we start with a Lie algebra $\mathfrak{g}$ with its Cartan subalgebra $\mathfrak{h}$. 
We denote its root system by $\Phi$ and the set of positive and negative roots by $\Phi^+$ and $\Phi^-$, respectively.
We then choose a   subalgebra $\subalg$ of $\mathfrak{g}$ containing $\mathfrak{h}$ and denote its root system by $\Phi_\subalg \subset \Phi  $.
Introduce a highest weight state $|\lambda\rangle $ of $\mathfrak{g}$,
\be
H^i|\lambda\rangle=\l^i |\lambda\rangle\ , \qquad
\forall H^i\in \mathfrak{h}\ , \qquad \qquad \qquad
E_\alpha|\lambda\rangle=0\ , \qquad
\forall \alpha\in \Phi^+\ .
\ee
Since $|\lambda\rangle $ is also an highest weight state of $\subalg$, we can use it to construct a finite dimensional irreducible representation $L_\subalg(\l)$ of $\subalg$.
Finally, the parabolic Verma module $M_{\subalg}(\l)$ is constructed by acting freely on $L_\subalg(\l)$ with the generators $E_{-\alpha}$ for all $\alpha \in \Psi^+ \equiv \Phi^+/\Phi_\subalg^+ $.
 
The case of conformal field theories in $d$ dimensions corresponds to
\be
\mathfrak{g}=\so(d+2)\ , \qquad\qquad \subalg=\so(2) \bigoplus  \so(d) \ .
\ee
 In fact a primary state $|\Ocal_\l \rangle$ can be labeled by $\l=(-\D,l_1,\dots,l_{[\frac{d}{2}]})$, where $-\D$ is the eigenvalue of $\so(2)$ and the set $l_1,\dots,l_{[\frac{d}{2}]}$ defines an irreducible representation of $\so(d)$.
In other words, the primary state transforms in an  irreducible representation $L_\subalg(\l)$ of the algebra $\subalg$. The set $\Psi^+$ contains the positive roots associated with the generators  $K_\mu$.
The parabolic Verma module is therefore obtained by acting with the negative roots $P_\m$   on the primary state, which gives the usual conformal multiplet
\be
M_{\subalg}(\l)=\{ |\Ocal_\l \rangle, P_\m |\Ocal_\l \rangle, P_\m P_\n |\Ocal_\l \rangle ,  \dots \} \ .
\ee
 
A parabolic Verma module is said to be simple if it does not contain any   submodule. Simplicity is equivalent to the absence of primary descendants in the conformal case. In order to state the conditions for simplicity of a parabolic Verma module we need to introduce some important concepts.
Firstly, we introduce the formal character of the parabolic modules $M_{\subalg}(\l)$ in terms of standard Verma modules $M(\l)$ and of the irreducible module $L_\subalg(\l)$
\begin{align}
{\rm ch}\, M_{\subalg}(\l)&=\sum_{w \in W_\subalg} (-1)^{\ell(w)} {\rm ch} \, M(w\cdot \l)
\label{eq:chMIintermsofchM} \\
&=\frac{{\rm ch}\, L_\subalg(\l)}{\prod_{\a \in \Psi^+} \left(1-e^{-\a} \right)} \ ,
\label{eq:characterMI}
\end{align}
where $W_\subalg$ is the Weyl group of (the semisimple part of) $\subalg$, $\ell(w)$ is the length of the Weyl reflection $w$ and the dot action of a Weyl reflection $w=s_\beta$ is defined by
\be
s_\b \cdot \l \equiv \l -\frac{2\langle \l+\r,\b\rangle}{\langle \b,\b \rangle}\b\ ,\qquad \qquad
\r=\frac{1}{2}\sum_{\a \in \Phi^+} \a\ .
\ee
We also define 
\be
\Psi^+_\l \equiv \left\{ 
\beta \in \Psi^+ :n_\b\equiv\frac{2\langle \l+\r,\b\rangle}{\langle \b,\b \rangle} \in \mathbb{Z}_{>0}
\right\}\ 
\ee
as the subset of $\Psi^+$ such that the $s_\beta \cdot \l =\l -n_\beta \beta$ for a non-negative integer $n_\beta$.
We are now ready to state Jantzen's simplicity criterion \cite{Jantzen}: $M_{\subalg}(\l)$ is simple if and only if 
\be
\sum_{\beta \in \Psi^+_\l}  {\rm ch}\, M_{\subalg}(s_\beta \cdot \l)
=0\ .
\label{eq:Jantzen}
\ee
Notice that when $\Psi^+_\l$ is empty, then the Jantzen\rq{}s criterion is trivially satisfied. Moreover, when $\Psi^+_\l$ is not empty, there are very simple geometrical ways to see if (\ref{eq:Jantzen}) is satisfied. To do so, it is very convenient to introduce the notion of a wall  between Weyl chambers. The wall $\Omega_\g$ is the hyperplane perpendicular to the root $\g \in \Phi$ that contains the point $-\rho$,
\be
\Omega_\g= \{ \l: \langle \l + \r, \g \rangle=0\} \ .
\ee
We can already dramatically simplify the search of simple modules using the following statement: when $\Psi^+_\l$ is not empty $M_{\subalg}(\l)$ can be simple only if $\l$ lives in a wall of a Weyl chamber. This means that we have to check condition (\ref{eq:Jantzen}) only when $\l \in \Omega_\g$ for some $\g$.

We now want to give some intuition on the reason why the sum in (\ref{eq:Jantzen}) can be zero. The main idea is that if two weights $\l$ and $\tilde \l$ are related by a Weyl transformation $w \in W_\subalg$ with odd length, then the sum of their characters is zero,
\be
w\cdot \l =\tilde{\l}\ ,\ \ \  w\in W_\subalg\ , \ \ \  (-1)^{\ell(w)}=-1\qquad  \Longrightarrow  \qquad  {\rm ch}\, M_{\subalg}(\l)+ {\rm ch}\, M_{\subalg}(\tilde{\l})=0\ .
\ee
Furthermore, if $w\cdot \l=\l$ then $ {\rm ch}\, M_{\subalg}(\l)=0$. 

When the parabolic module $M_{\subalg}(\l)$ is not simple, it will contain at least one primary descendant. Notice that the weight vectors  $ s_\b\cdot \l=\l-n_\b \b $ with $\beta \in \Psi^+_\l$ are good candidates to be primary descendants because their quadratic  Casimir is equal to the one of $\l$,
\be
\mathcal{C} |s_\b\cdot \l \rangle=(s_\b\cdot \l,s_\b\cdot \l+2\r)=(\l,\l+2\r) =\mathcal{C} | \l \rangle \  .
\ee
The general decomposition of the module $M_{\subalg}(\l)$ into irreducible  modules is rather complicated, however  this question can be approached using the   \emph{Kazhdan-Lusztig theory} \cite{KazhdanLusztig}.\footnote{See appendix C of \cite{arXiv:1312.5344} (and references therein) for a more physics oriented summary.}

The Kazhdan-Lusztig conjecture for parabolic Verma modules (\emph{relative Kazhdan-Lustzig conjecture}) \cite{Deodhar,CasianCollingwood}
tells us how to decompose a parabolic 
Verma module $M_{\subalg}(\l)$ into irreducible modules $L(\l)$. 

For this purpose, it is convenient to parametrize the weights $\l$ and $\m$ by elements of the Weyl group.
It is known that we can parametrize the weight of independent irreducible modules
by an element of the coset: $w\in W_{\mathfrak{p}}\backslash W$.
Let us write the corresponding weight by $\lambda(w)$\footnote{See e.g.\ \cite{CasianCollingwood} for an explicit description of $\lambda(w)$.}.
Then, the characters of the irreducible modules are given by appropriate linear combinations of the characters of parabolic Verma modules, 
\be
{\rm ch} \, L( \lambda(x))=
\sum_{x\le w}  m_{\lambda(x), \lambda(w)}  \,
{\rm ch} \, M_{\subalg}(\lambda(w)) \ ,
\label{eq:KLconjecture}
\ee
where $\le$ in the sum denotes the Bruhat ordering.
The relative Kazhdan-Lusztig (KL) conjecture states that
the multiplicity $m_{\lambda(w), \lambda(x)}$ is a special value 
of the Kazhdan-Lusztig polynomial $P_{x,w}(q)$:
\be
m_{\lambda(w), \lambda(x)}=(-1)^{\ell(w)-\ell(x)} P_{x,w}(1)  
\ .
\label{m_P_KL}
\ee
In appendix \ref{app.KLconjecture} we present explicit expressions for the KL polynomial.


In the following we apply both the Jantzen\rq{}s criterion and \emph{relative Kazhdan-Lusztig conjecture} to conformal field theories. We will first study the example of a CFT$_3$ and then we will generalize it to CFT in generic odd and even dimensions. A summary of the results is presented in section \ref{sec.Connections}, where we will completely classify the poles of conformal blocks and their residues.

\subsection{Example: \texorpdfstring{$\so(5)$}{so(5)}}
\label{Section:CFT3}

\begin{figure}
\graphicspath{{Fig/}}
\def\svgwidth{5.8 cm} \centering
 \input{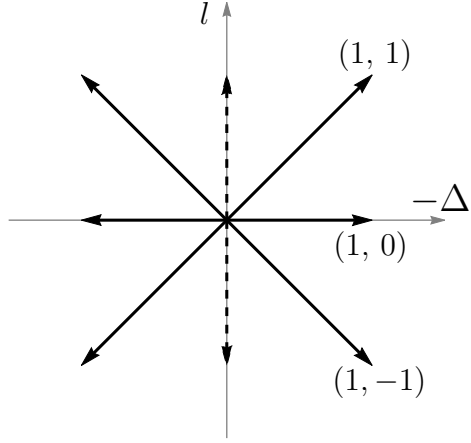}
\caption{\label{fig:SO5roots} 
The root system of $\so(5)$. The dashed roots belong to $\so(3)\subset \so(5)$.
}
\end{figure}

It is instructive to apply the general statement of the previous section to the case of 3 dimensional CFTs. In this case, $\subalg=\so(2)\bigoplus \so(3)$, thus we can label the highest weight by $\l=(-\Delta,l)$. The finite dimensional representation $L_\subalg(\l)$ are the spin $l$ representations of $\so(3)$ with eigenvalue $-\Delta$ under $\so(2)$,
\be
{\rm ch}\, L_\subalg(\l)=\frac{
e^{(-\D,l)}-e^{(-\D,-l-1)}}
{1-e^{-(0,1)} } =
\sum_{s=-l}^l e^{(-\D,s)}\ .
\ee
The root system of $\so(5)$, depicted in figure \ref{fig:SO5roots}, leads to
\be
\Psi^+=\left\{ 
(1,1),(1,0),(1,-1)
\right\}\ .
\ee
From the numbers
\be
n_{(1,1)}=l+2-\D\ ,\qquad
n_{(1,0)}=3-2\D \ ,\qquad
n_{(1,-1)}=1-l-\D \ ,
\label{numbersCFT3}
\ee
we conclude that for  $\Delta>l+1$ the set $\Psi_\l^+$ is empty and therefore the module $M_{\subalg}(\l)$ is simple. 
However, if any of the numbers in (\ref{numbersCFT3})
takes a positive integer value, then we have to check  condition (\ref{eq:Jantzen}). 

\begin{figure}
\graphicspath{{Fig/}}
\def\svgwidth{\columnwidth}
 \input{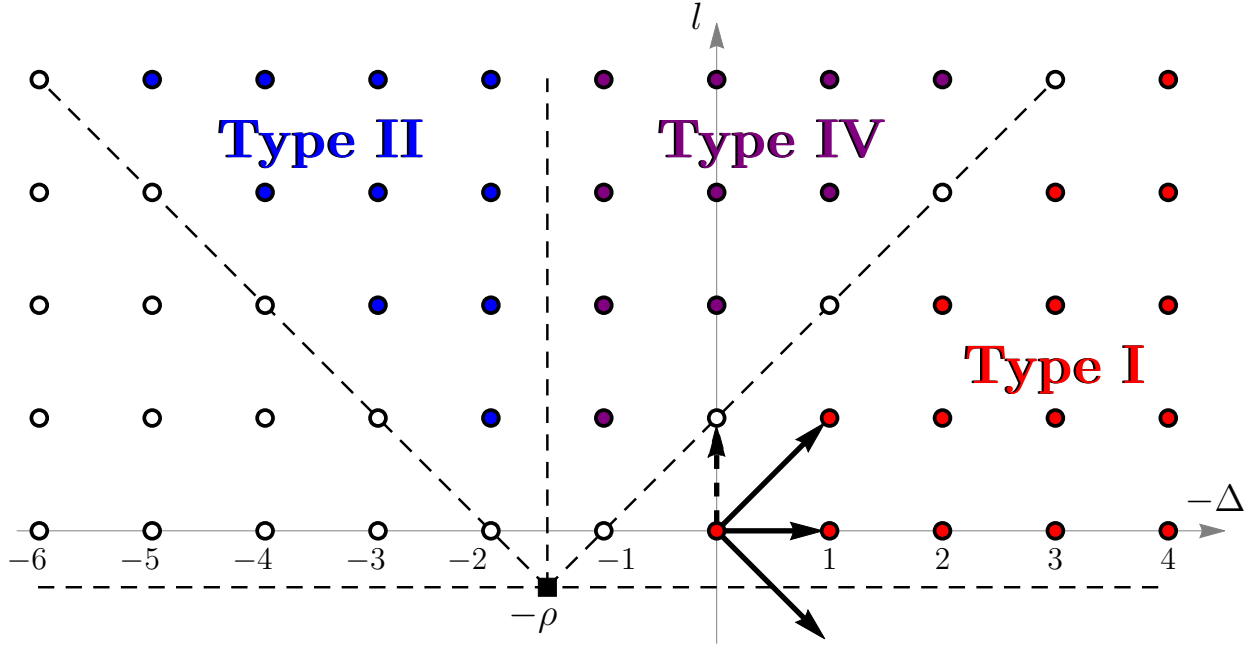}
\caption{\label{fig:WeightSpaceCFT3} 
\ Integral weights of $\so(5)$.}
\end{figure}

First, we notice that $n_\b$ in (\ref{numbersCFT3}) is an integer only when $\D$ is either an integer or a semi-integer.
When $\D$ is an integer there exist three special cases for which  $\Psi_\l^+$ is not empty and the module is not simple: \footnote{The denomination of the types is made to match section \ref{subsec.nullstates}. We will comment later on the type \IV.}
\be \nonumber
\begin{array}{lll}
\textrm{Type}\; \IC. &\quad  \Delta=1-l-k \, ,\; (k=1, 2,\dots) &\quad \Longrightarrow \quad \Psi^+_\l=\{(1,1),(1,0),(1,-1)\}  \ , \\
\textrm{Type}\; \IIA.& \quad  \Delta=l+2-k \, ,\; (k=1,2,\dots,l)        &\quad \Longrightarrow \quad \Psi^+_\l=\{(1,1)\}  \ , \\
\textrm{Type}\; \IVB. &\quad  \Delta= 2-k \, ,\; \ \ \ \ \, (k= 1, 2,\dots,l) &\quad \Longrightarrow \quad \Psi^+_\l=\{(1,1),(1,0)\} \ . \\
\end{array}
\ee
This can be understood geometrically from figure \ref{fig:WeightSpaceCFT3}.
The three special regions \IC, \IIA and \IVB are just the colored dots in the Weyl chambers. The dot action $s_\b \cdot \l$ corresponds to reflections (towards the left) on the dashed lines. Moreover, the set $\Psi^+_\l$ is simply the set of allowed dot action reflections. It is then trivial to see why in case \IIA there is only one element in $\Psi^+_\l$, while in the cases \IVB and \IC there are respectively $2$ and $3$.

There exists one special case in which $\Psi^+_\l$ is not empty but the module is simple. This happens when $\l$ belongs to the wall $\Omega_{(1,-1)}$ defined by $\Delta=1-l$. Notice that the wall  $\Omega_{(1,-1)}$ is given by the line perpendicular to $(1,-1)$ that passes trough $-\r$, which corresponds to the dashed line between the regions \IVB and \IC in figure \ref{fig:WeightSpaceCFT3}. 
It easy to check that $ \Psi^+_\l=\{(1,1),(1,0)\}$ and 
\be
 {\rm ch}\, M_{\subalg}(s_{(1,1)}\cdot \l )+
 {\rm ch}\, M_{\subalg}(s_{(1,0)}\cdot \l )=0\ ,
\ee
thus the condition (\ref{eq:Jantzen}) holds and the module is simple.

When the module is not simple we can understand the structure of submodules using Kazhdan-Luzstig theory. In this case, all Kazhdan-Luzstig polynomials are equal to 1 and using (\ref{eq:chMIintermsofchM}) and (\ref{eq:KLconjecture}) we obtain (see appendix \ref{app.KLconjecture})
\begin{eqnarray}
 \label{KLso(5):C}
\textrm{Type}\; \IC.\quad &{\rm ch}\,M_{\subalg}(\l) &= {\rm ch}\, L (\l) 
+{\rm ch}\, L (s_{(1,-1)}\cdot \l) \ , \\
 \label{KLso(5):A}
\textrm{Type}\; \IIA.\quad &{\rm ch}\,M_{\subalg}(\l) &= {\rm ch}\, L (\l) 
+{\rm ch}\, M_{\subalg} (s_{(1,1)}\cdot \l) \ , \\
\label{KLso(5):B}
\textrm{Type}\; \IVB.\quad&{\rm ch}\,M_{\subalg}(\l) &= {\rm ch}\, L (\l) 
+{\rm ch}\, L (s_{(1,0)}\cdot \l) \ .
\end{eqnarray}
In the case \IIA the module $M_{\subalg}(\l)$ contains the simple parabolic submodule $M_{\subalg} (s_{\b}\cdot \l)$, where $\b$ is the unique element in $\Psi^+_\l$. Geometrically we can check that when $\l$ is a blue dot in figure \ref{fig:WeightSpaceCFT3}, $s_{(1,1)} \cdot \l$ is a white dot which correspond to a simple parabolic module.
In the cases \IVB and \IC, things are more interesting and we find that the module $M_{\subalg}(\l)$ does not contain any parabolic submodule, instead it decomposes into the two irreducible modules $L(\lambda) $ and $ L(s_{\b} \cdot \lambda)$ where $\b$ is a root in $\Psi^+_\l$. Notice that the actual root $\b$ that we found in (\ref{KLso(5):C}) and (\ref{KLso(5):B}) is the one corresponding to the lowest $n_\beta>0$.
Geometrically  this means that $s_\b$ is the Weyl transformation that maps $\l$ to the closest Weyl chamber on the left.

\begin{figure}
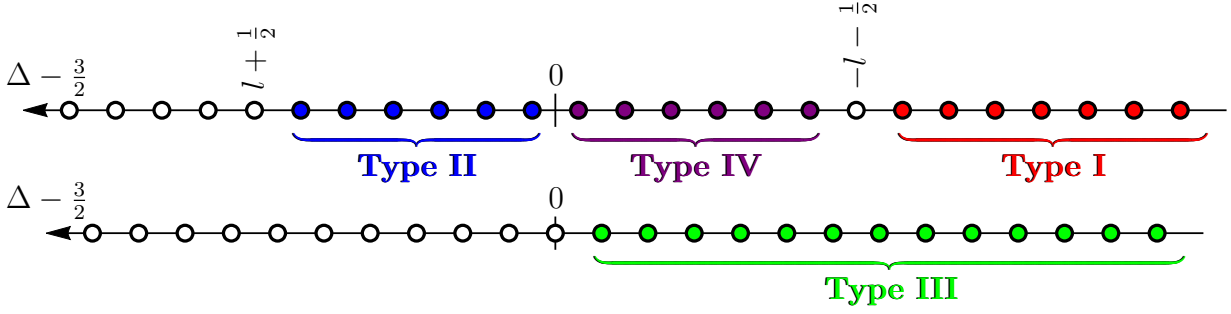

\centering
\graphicspath{{Fig/}}
\def\svgwidth{16 cm}
 \input{halfIntegersCFT3.pdf_tex}
\def\svgwidth{16 cm}
 \input{IntegersCFT3.pdf_tex}
\caption{\label{fig.weightsSO5} Summary of the weights of $\so(5)$, for $\D$ integer (above) and $\D$ half-integer (below).}
\end{figure}

If $\D$ is a half-integer there is only a single special case when  $\Psi_\l^+$ is not empty:
\begin{align}
&\textrm{Type}\; \IIID. \quad  \Delta=\frac{3}{2}-k \qquad (k=1,2,\dots)        &  \Longrightarrow \quad \Psi^+_\l=\{ (1,0)\} \ .
\end{align}
In the case \IIID, as in the case \IIA, there exists only one element in $\Psi^+_\l$. Therefore the module $M_{\subalg}(\l)$ contains only the simple  parabolic submodule $M_{\subalg} (s_{(1,0)}\cdot \l)$
\be
\textrm{Type}\; \IIID. \quad {\rm ch}\,M_{\subalg}(\l) = {\rm ch}\,L (\l)  \label{KLso(5):D}
+{\rm ch}\, M_{\subalg} (s_{(1,0)}\cdot \l) \ .
\ee
The full set of weights is summarized  in figure \ref{fig.weightsSO5}.

The structure of submodules has important implications for the analytic structure of conformal blocks as functions of $\Delta$.
The absence of second generation primary descendants leads to absence of higher order poles. Therefore, we conclude that in 3D the conformal blocks only have single poles. 
Moreover, the residues of the poles are given by the blocks associated to the submodules that become null. This means that for poles of type \IVB{}   and \IC the residues are not conformal blocks but blocks associated to the exchange of the irreducible modules $L(\l)$.
These can be defined at specific values of $\D$ as the solutions to the Casimir differential equation that start with the same leading OPE as the conformal block at that dimension but have zero coefficient associated with the exchange of all other primary descendants. 
This new kind of blocks was not important in the previous sections because we only considered operators that can appear in the OPE of two scalar primary operators.

In section \ref{subsec.nullstates} we found the full set of primary descendants which can appear in  the OPE of two scalars.
Each of the three types corresponds to one case in which the parabolic module is not simple.
The case $\IVB$ corresponds instead to a new type defined by
\be
|\D+2n-1,l\, ;z\rangle   =\Dcal_{\IV,n}|\D,l\, ;z \rangle \equiv \e(P,z,D_z)  \calW_0 \cdot  \calW_1 \cdots \calW_{n-2}  
|\D,l\, ;z \rangle  \ , \label{stateIV}
\ee
with $n=1\dots l$ and where $\e(P,z,D_z)$ stands for the contraction of the 3-dimensional $\epsilon$-tensor with the vectors $P^\mu$, $z^\nu$ and $D_z^\a$ and 
\be
 \calW_j=  P^2-2\frac{ (P\cdot z) (P\cdot D_z)}{(l+j+1) (l-j-1)} \ .
\ee
In section \ref{subsec.nullstates} we did not include the type {\IV} states, since $\Dcal_{\IV,n}$ changes the parity of the primary, thus  the state $|\D+2n-1,l\,;z\rangle$ is not present in the OPE of two scalars.

It is important to stress that the operators $\Dcal_{A}$ simply implement Weyl reflections. For example $\Dcal_{\I,n}$ reflect the chamber $\IC$ into the chamber $\IVB$, while $\Dcal_{\IV,n}$ reflects $\IVB$ in $\IIA$ and $\Dcal_{\II,n}$ maps $\IIA$ into the Weyl chamber on the left of figure \ref{fig:WeightSpaceCFT3} which has simple module. 
The operator  $\Dcal_{\III,n}$, like $\Dcal_{\IV,n}$, is a Weyl reflection with respect to the root $(1,0)$.

Since we have the full set of operators which describe the three dimensional case we are able to check the consistency of the formulas (\ref{KLso(5):A},\ref{KLso(5):B},\ref{KLso(5):C},\ref{KLso(5):D}). These formulas imply that  the module is simple after we apply one Weyl reflection. Therefore if we apply two consecutive operators $\Dcal_{A}$ in such a way to implement a double Weyl reflection we need to obtain zero. In fact we checked in many cases that 
\begin{align}
 \Dcal_{\IV,l+1} \ \Dcal_{\I,n} \ \,
 |\Delta=1-l-n,l\, \rangle &= 0 &
 (n=1,2,\dots ) \ ,\\
 \Dcal_{\II, 1+l-n}\ \Dcal_{\IV,n} \ \ 
 |\Delta=2-n,l\, \rangle &= 0 &  (n=1, \dots,l)\ ,
\end{align}
hold. This is an independent check of the absence of nested second generation primary descendant which would give rise to multiple poles in the conformal blocks.

\subsection{Odd Spacetime Dimension}
\label{Section:Odddim}

We now consider the general case in odd spacetime dimension $d=2N+1$. 
The complexified conformal algebra is $\so(2N+3)$. The set of positive roots is   
\be
\Phi^+ =\{ \a_{jk}^{+ -},\a_{jk}^{+ +},\a_{j}^{+}\}\ ,
\ee
 defined in (\ref{RootsSO2N+1}), and the vector $\rho$ is given by
\be
 \r =\left(N+\frac{1}{2},N-\frac{1}{2},\dots,
 \frac{3}{2},\frac{1}{2}\right) \ .
\ee
 We consider the highest weight  $\l=(-\D, l_1,\dots ,l_N )$ and the parabolic Verma modules based on the subalgebra $\subalg=\so(2)\bigoplus \so(2N+1)\subset \so(2N+3)$. We have 
\be
 \Psi^+ = \{ \alpha_0^+,\alpha_{01}^{++},\alpha_{02}^{++},\dots,\alpha_{0N}^{++},\alpha_{01}^{+-},\alpha_{02}^{+-}, \dots,\alpha_{0N}^{+-} \} \ ,
 \ee 
with
\begin{align}
n_{\alpha_0^+}\; &=2N+1-2\Delta  \ , &
\label{nalpha0+}\\
n_{\alpha_{0j}^{++}}&=2N+1-\Delta+l_j-j   &(j=1,2,\dots,N) \ ,\\
n_{\alpha_{0j}^{+-}}&= -\Delta-l_j+j  &(j=1,2,\dots,N) \ .
\label{nalpha0j+-}
\end{align}
We conclude that for generic real values of $\Delta$ the module $M_{\subalg}(\l)$ is simple. For integer values of $\Delta$ the question is much more non-trivial.  It is clear that for $\Delta \ge 2N+l_1$ the module $M_{\subalg}(\l)$ is simple. For smaller integer values of $\Delta$ there is always some $n$ from (\ref{nalpha0+}-\ref{nalpha0j+-}) which takes a positive integer value. In these cases we have to check condition (\ref{eq:Jantzen}). The results for integers roots are summarized in figure \ref{fig:DeltaLineSO(odd)}. The semi integer case is similar to   the case \IIID in $\so(5)$.

\begin{figure}
\graphicspath{{Fig/}}
\def\svgwidth{\columnwidth}
 \input{RecRelFig2.pdf_tex}
\caption{\label{fig:DeltaLineSO(odd)} 
The diagram represents all  integer values of $\Delta $ when $d=2N+1$. The white dots mark values of $\Delta$ for which the module $M_{\subalg}(\l)=M_{\subalg}(-\Delta,l_1,\dots,l_N)$ is simple. The coloured dots  mark the values for which $M_{\subalg}(\l)$ is not simple. In the set $\IIA_k$, there is a submodule with smaller $l_k$ and all other $l$'s unchanged. In the set $\IVB$, there is a submodule with exactly the same $l$'s. In the set $\IC_k$ there is a submodule with larger $l_k$ and all other $l$'s unchanged.
}
\end{figure}

To explain figure \ref{fig:DeltaLineSO(odd)} we start by studying the special integer values of $\Delta$ for which $M_{\subalg}(\l)$ is simple.
These organize in two sequences.
The first is when $\l$ belongs to the wall of the Weyl chamber $\Omega_{\a^{++}_{0k}}$, given by  
\be
\Delta=2N+1+l_k-k \qquad (k=1,\dots,N) \ .
\ee
The second is when $\l$ belongs to the wall $\Omega_{\a^{-+}_{0k}}$,  
\be
\Delta=k-l_k  \qquad  (k=1,2,\dots,N) \ .
\ee
In both cases, Jantzen's criterion implies that the module $M_{\subalg}(\l)$ is simple. We explain this in detail in appendix \ref{App:Jantzen}.

Let us now describe the integer values of $\D$ for which $M_{\subalg}(\l)$ is not simple. These cases are naturally divided into four types 
\begin{align}
\l^\star_{\IC_k, n}&\equiv\{ \l :\Delta=k-l_k-n \}   \qquad   
&&(n=1,2,\dots,l_{k-1}-l_k)  \ , \\
\l^\star_{\IIA_k, n}&\equiv\{ \l : \Delta=2N+1+l_k-k-n \} \qquad
&&  (n=1,2,\dots,l_{k}-l_{k+1} )  \ , \\
\l^\star_{\IIID, n} &\equiv\{ \l : \Delta=\tfrac{1}{2}+N-n\} \qquad &&(n=1,2,\dots) \ , \\
\l^\star_{\IVB, n}&\equiv\{ \l : \Delta=N+1-n \} \qquad&    &(n=1,2,\dots,l_{k} ) \ ,
\end{align}
where $k=1,2,\dots,N$ and we defined $l_{N+1}\equiv0$ and $l_{0}\equiv\infty$. Each of these cases has  the following set of roots $\Psi^+_\l$ 
\be
\Psi^+_\l= 
\left\{
\begin{array}{l l}
\{\alpha_{01}^{++},\dots,\alpha_{0N}^{++},\a_0^+,\alpha_{0N}^{+-},\dots,\alpha_{0k}^{+-}\}  \qquad & (\l=\l^\star_{\IC_k, n}) \ , \nonumber \\
\{\alpha_{01}^{++},\dots,\alpha_{0k}^{++}\}  \qquad &(\l=\l^\star_{\IIA_k, n}) \ ,\nonumber \\
\{ \a_0^+ \} \qquad &  (\l=\l^\star_{\IIID, n}) \ , \nonumber \\
 \{\alpha_{01}^{++},\dots,\alpha_{0N}^{++},\a_0^+\} \qquad & (\l=\l^\star_{\IVB, n})\ . \nonumber
\end{array}
\right.
\ee
 In the cases $\IIA_1$ and $\IIID$ the set $\Psi^+_\l$ contains only one element and the decomposition in submodules is straightforward. In all the other cases, one can work out the detailed decomposition of the parabolic modules using Kazhdan-Lusztig theory as we show in appendix  \ref{app.KLconjecture}. The result is as follows
\be
\begin{array}{lc}
{\rm ch}\, M_{\subalg}(\l^\star_A) ={\rm ch}\, L(\l^\star_A)+{\rm ch}\, M_{\subalg}(\l_A) \ ,\qquad  &  A= (\IIA_1,n), (\IIID,n) \ , \\
{\rm ch}\, M_{\subalg}(\l^\star_A) ={\rm ch}\, L(\l^\star_A)+{\rm ch}\, L(\l_A) \ ,  \qquad & \mbox{otherwise} \ ,
\end{array}
\ee
where the irreducible submodules are labeled by the following weights $\l_A$
\be \label{newmodules}
\begin{array}{llcl}
 \quad  \l_{\IC_k , n}&\equiv s_{\alpha_{0k}^{+-}} \cdot \l^\star_{\IC_k, n}&=&(l_k-k,l_1,\dots,l_k+n
,\dots,l_N)
 ,  \\
  \quad \l_{\IIA_k , n}&\equiv s_{\alpha_{0k}^{++}} \cdot \l^\star_{\IIA_k, n} &=&(-2N-1-l_k+k,l_1,\dots,l_k-n 
,\dots,l_N)
 \ ,  \\
\quad   \l_{\IIID , n}&\equiv s_{\alpha_{0\phantom{k}}^{+\phantom{+}}} \cdot \l^\star_{\IIID, n} \phantom{_{_{k}}}&=&
 \left(\tfrac{1}{2}+N+q,l_1,\dots  ,l_N\right)
\ ,  \\
 \quad  \l_{\IVB , n}&\equiv s_{\alpha_{0\phantom{k}}^{+\phantom{+}}} \cdot \l^\star_{\IVB, n} \phantom{_{_{k}}}&=&(1-N+q,l_1,\dots,l_N) 
 \ . 
\end{array}
\ee
Notice that in type $\IIA_k$ the first submodule has smaller $l_k$, in type \IVB and \IIID  the first submodule has the same $l_k$'s, and in type $\IC_k$ the first submodule has larger $l_k$. 

\subsection{Even Spacetime Dimension}
\label{Section:Evendim}

\begin{figure}[t!]
\graphicspath{{Fig/}}  
\def\svgwidth{\columnwidth}
 \input{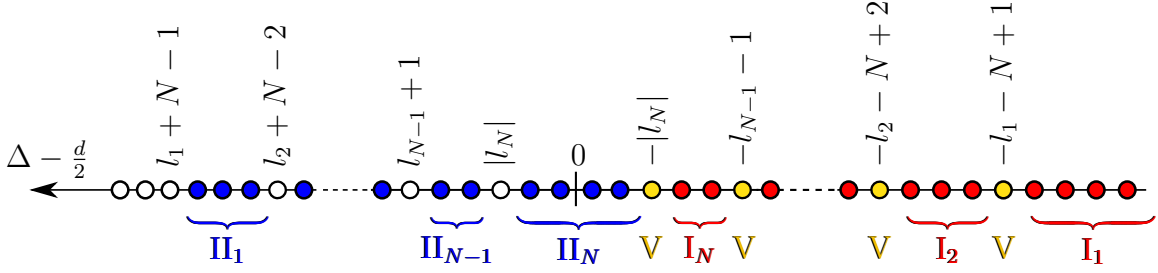}
\caption{\label{fig:DeltaLineSO(even)} 
The diagram represents all  integer values of $\Delta $ when $d=2N$. The white circles mark values of $\Delta$ for which the module $M_{\subalg}(\l)=M_{\subalg}(-\Delta,l_1,\dots,l_N)$ is simple. The coloured dots  mark the values for which $M_{\subalg}(\l)$ is not simple. In the set $\IIA_k$, there is a submodule with smaller $|l_k|$ and all other $l$'s unchanged.  In the set $\IC_k$ there is a submodule with larger $|l_k|$ and all other $l$'s unchanged.
In case \VE (yellow circles), there is a submodule with the opposite sign of $l_N$ and all other $l$'s unchanged.}
\end{figure}

The case $d=2N$ is similar.
The set of positive roots is   
\be
\Phi^+ =\{ \a_{jk}^{+ -},\a_{jk}^{+ +} \}\ ,
\ee 
and the vector $\rho$ is given by
\be
 \r =\left(N ,N-1,\dots,
 1,0\right) \ .
\ee 
 We have 
\be
 \Psi^+ = \{  \alpha_{01}^{++},\alpha_{02}^{++},\dots,\alpha_{0N}^{++},\alpha_{01}^{+-},\alpha_{02}^{+-}, \dots,\alpha_{0N}^{+-} \}
 \ee 
with
\begin{align} 
n_{\alpha_{0j}^{++}}&=2N -\Delta+l_j-j  &(j=1,2,\dots,N) \ , \\
n_{\alpha_{0j}^{+-}}&= -\Delta-l_j+j   &(j=1,2,\dots,N) \ . 
\end{align}
We conclude that for generic values of $\Delta$ the module $M_{\subalg}(\l)$ is simple. For integer values of $\Delta$ the answer is summarized in figure \ref{fig:DeltaLineSO(even)}. 
For $\Delta \ge 2N+l_1-1 $ the set $\Psi_\l^+$ is empty and the module is simple. For the special values 
\be
\Delta=2N+|l_k|-k  \qquad (k=1,2,\dots,N) \ ,
\ee
the set $\Psi_\l^+$ is not empty but one can show that the modules are still simple (see appendix \ref{App:Jantzen}).
For all other integer values of $\D$ the module $M_{\subalg}(\l)$ is not simple.
We divide these cases into three groups depending on the first submodule that appears in the decomposition of $M_{\subalg}(\l)$. Cases $\IIA_k$ and $\IC_k$ are very similar to what happens in odd dimensions, in the sense that they correspond to decreasing and increasing $|l_k|$.
In case \VE, the first submodule flips the sign of $l_N$ and keeps all other $l$'s unchanged.

In appendix \ref{app.KLconjecture}, we use Kazhdan-Lusztig theory to find the precise decomposition of the parabolic modules  into irreducible modules. We show the non-trivial decompositions that arise in even spacetime dimension which implies a more complicated pole structure of the conformal blocks.
\subsection{Comments on Conformal Representation Theory} 
\label{sec.Connections}
It is useful to summarize the results that we obtained from the study of conformal representation theory stressing the implications for the study of conformal blocks and unitary CFTs. 

First we saw that the odd dimensional case is easier than the even dimensional one. Thus the analytic  continuation of the conformal blocks in the spacetime dimension $d$ should be easier starting from odd dimension.
In this case, we know the full set of poles of any conformal block:
\be
\begin{array}{lcll}
\Delta^\star_{\IC_k,n }&=&k-l_k-n\qquad & (n=1,2,\dots,l_{k-1}-l_k) \ ,\\
\Delta^\star_{\IIA_k,n }&=&d+l_k-k-n  \qquad  & (n=1,2,\dots,l_k -l_{k+1})\ , \\
\Delta^\star_{\IIID,n}&=&\frac{d}{2}-n \qquad &  (n=1,2,\dots)\ , \\ 
\Delta^\star_{\IVB,n}&=&\frac{d}{2}+\frac{1}{2}-n \qquad & (n=1,2,\dots,l_{k})\ .\\
\end{array}
\ee
We conjecture that a conformal block $G_{\l}$ associated to the exchange of the operator $\Ocal_\l$, labeled by an highest weight $\l=(-\D,l_1,\dots,l_N)$  of $\so(d+2)$, has the following behavior at the pole
\be \label{genericCBpole}
G_{\l}^{(p,q)} \underset{\D\rightarrow  \D^\star_A}{\longrightarrow} \sum_{p\rq{},q\rq{}} \frac{(R_A)_{p p\rq{} q q\rq{}}}{\D- \D^\star_A}
G_{\l_A}^{(p\rq{},q\rq{})} \ ,
\ee 
where $A=({\IC_k,n }),(\IIA_k,n ),({\IIID,n }),({\IVB,n })$, $\l_A$ are the ones defined in 
(\ref{newmodules}) and the coefficients $(R_A)_{p p\rq{} q q\rq{}}$ can be computed using the techniques introduced in the previous sections. The regular part in $\D$ of the conformal blocks can be also obtained from a direct computation (see appendix \ref{App:CBatLargeDelta}).
Combining the knowledge of the pole structure \eqref{genericCBpole} and the regular part one can in principle build any conformal block.

At a first sight formula \eqref{genericCBpole} may look divergent since  $G_{\l_A}^{(p\rq{},q\rq{})}$ has a pole precisely at the values of $\D$ specified by $\l_A$ (for all $A$ which are not of the type $(\II_1,n)$ and $(\III,n)$). However in \cite{Costa:2016xah}, for the case of two external vectors, it is shown that the specific linear combination provided by the coefficients $R_A$ is such that the divergence is canceled. We therefore conjecture that \eqref{genericCBpole} is correct.

From representation theory we also obtain a very sharp way to constrain the spectrum of unitary theories. Since unitary theories can only contain states $| \Ocal_\l \rangle $ with positive norms, then all the respective modules $M_{\subalg}(\l )$ have to be simple.  Watching figure \ref{fig:DeltaLineSO(odd)} it is easy to understand that this requires that the value of the conformal dimension $\D$ of any operator has to satisfy
\begin{align}
&\D\ge \D^\star_A \qquad  \mbox{for any } A  \label{UnitarityComplete} \ .  
\end{align}
Actually (\ref{UnitarityComplete}) gives many redundant conditions, for example it is clear that the bounds are optimized when $n$ is as low as possible.
In fact when the operator is a scalar we obtain that  the bound (\ref{UnitarityComplete}) reduce to the usual
\be
 \D \ge \frac{d-2}{2} \ .
\ee
Moreover for a generic  operator with $l_1>l_2>0$ we have
\begin{align} \label{ConsCurrentUnitarity}
&\D \ge l_1+d-2  \ ,
\end{align}
which gives the usual unitarity bound for tensor operators. When the first $k$ lines of the Young tableau are of the same length, the bound becomes 
\be
\D \ge d-1-k+l  \qquad (l_1,\dots,l_k=l>0) \ .
\ee
This analysis matches the results known in the literature  \cite{Mack:1975je, Metsaev:1995re, Minwalla:1997ka}.

\section{Conclusion} 
In the first part of this paper we explained how to build the conformal blocks studying their analytic behavior in terms of the conformal dimension of the exchanged operator. Both for pedagogical reasons and to test the method, we considered simple cases in which the external operators are:   four scalars;
 tree scalars and one vector;
tree scalars and one conserved current.

We believe that this method can be useful to obtain conformal blocks for more generic cases.  For this reason in chapter \ref{StructureofCFT} we gave a complete classification of  the cases in which a conformal family becomes reducible, which amounts to classify all the possible poles $\D^\star_A$ of any  conformal block. Moreover in formula \eqref{genericCBpole} 
\be \nonumber
G_{\l}^{(p,q)} \underset{\D\rightarrow  \D^\star_A}{\longrightarrow} \sum_{p\rq{},q\rq{}} \frac{(R_A)_{p p\rq{} q q\rq{}}}{\D- \D^\star_A}
G_{\l_A}^{(p\rq{},q\rq{})} \ ,
\ee 
we conjecture that the residues at the poles are always proportional to conformal blocks labeled by new representations $\l_A$, which are known and summarized in formula \eqref{newmodules}. If \eqref{genericCBpole} is correct, this method would provide a new independent way to build any conformal block.

Equation \eqref{genericCBpole} is well motivated by representation theory but it looks divergent since  the conformal blocks $G_{\l}^{(p,q)}$ have poles in $\D$ for most of the $\l=\l_A$. However we conjecture that the divergence is fictitious and it is canceled by the specific linear combination provided by the coefficients $R_A$. An important extension of this work is given in \cite{Costa:2016xah} in which it is shown that the cancellation holds for  the conformal blocks of two external vectors and two scalars. It would be interesting to study more cases and prove that the residue in \eqref{genericCBpole} is indeed finite.

\section*{Acknowledgements} 

We would like to thank Miguel Costa, Tobias Hansen, James E.\ Humphreys, Yoshiki Oshima, Balt van Rees, Slava Rychkov and  David Simmons-Duffin  for stimulating discussions.
J.P.\ and M.Y.\ would like to thank KITP/UCSB for hospitality, where this work has been initiated
 during the program ``New Methods in Nonperturbative Quantum Field Theory''.
 The contents of this paper were presented by E.T.\footnote{``Back to the Bootstrap IV'' (July 2014, Univ.\ Porto), ``Bologna Workshop on CFT and Integrable Models'' (Sep.\ 2014), ICTP-SAIFR (Nov.\ 2014), IDPASC workshop (Univ.\ Porto, Mar.\ 2015).} and M.Y.\footnote{Caltech (Dec.\ 2015), KEK theory workshop (Jan. 2015), Univ.\ North Carolina (Mar.\ 2015), Univ.\ Chicago (Apr.\ 2015), IAS (May.\ 2015).}, and we thank the audience for feedback.
The
research leading to these results has received funding from the [European Union] Seventh
Framework Programme [FP7-People-2010-IRSES] and [FP7/2007-2013] under grant agreements
No 269217, 317089 and 247252, and from the grant CERN/FP/123599/2011.
Centro de 
Fisica 
do Porto is partially funded by the Foundation for Science and Technology
of Portugal (FCT).
E.T.\ would like to thank FAPESP grant 2011/11973-4 for funding his visit to ICTP-SAIFR from November 2014 to March 2015 where part of this work was done. He would also like to thank Perimeter Institute for for the hospitality in the summer of 2015. The work of E.T. is supported by the FCT fellowship SFRH/BD/51984/2012.
M.Y.\ is supported in part by the World Premier International Research Center
Initiative (MEXT, Japan), by  JSPS Program for Advancing Strategic
International Networks to Accelerate the Circulation of Talented Researchers,
by JSPS KAKENHI Grant Number 15K17634,
and by Institute for Advanced Study.
He would also like to thank Aspen Center for Physics (NSF Grant No.\ PHYS-1066293) 
and Simons Center for Geometry and Physics for hospitality, where part of this work has been performed.

\appendix

\section{Primary Descendant States}\label{app.norm}
In this appendix we aim to demonstrate that a descendant $\Ocal_A$ of a primary $\Ocal$ becomes a primary when $\D= \D^{\star}_{A}$. Moreover we provide more details on the computation of the norm of $|\Ocal_A\rangle$.

\subsection{Type \I}
We first want to demonstrate that a descendant of type \I becomes a primary when $\D= \D^{\star}_{\I,n}$, namely
\be
 (K\cdot z\rq{}) (P \cdot z)^n | \D, l \,;z \rangle = 0\ , \qquad\qquad \textrm{if }\D= \D^{\star}_{\I,n}\equiv 1-l-n\ . \label{PrimaryCondI}
\ee
Commuting $(K\cdot z\rq{})$ through all the $ (P \cdot z)$ we obtain
\begin{align}
[ (K\cdot z\rq{}), (P \cdot z)^n ] 
&=\sum_{j=1}^n  (P \cdot z)^{n-j}[(K\cdot z\rq{}),(P \cdot z)](P \cdot z)^{j-1} \nonumber\\ 
&= 2 i \sum_{j=1}^n  (P \cdot z)^{n-j}[(z \cdot z\rq{}) D - (z\rq{} \cdot J \cdot z)] (P \cdot z)^{j-1}\nonumber\\ 
&= 2 i \sum_{j=1}^n  (P \cdot z)^{n-1}[(z \cdot z\rq{}) (D+2 i (j-1)) - (z\rq{} \cdot J \cdot z)]\nonumber \\ 
&=2 i\, n \, (P \cdot z)^{n-1}  [(z \cdot z\rq{}) (D+ i (n-1)) - (z\rq{} \cdot J \cdot z)] \ . \label{commutatorKPzn}
\end{align}
Acting with (\ref{commutatorKPzn}) on a primary state we get
\be \label{PrimaryCondIcheck}
 (K\cdot z\rq{}) (P \cdot z)^n  | \D, l \,;z \rangle= -2 n(\D+n+l-1) (z \cdot z\rq{}) (P \cdot z)^{n-1}| \D, l \,;z \rangle \ ,
\ee
which proves \eqref{PrimaryCondI}.

We can then compute the norm $N_{\I, n}$ defined by
\be
 N_{\I, n} \equiv \frac{\langle \D +n , l-n ; z\rq{} | \D+n, l-n ; z \rangle}{(z \cdot z\rq{})^{n + l}}=\frac{\langle \D , l\, ;z\rq{} |(K\cdot z\rq{})^n (P \cdot z)^n| \D, l\, ; z \rangle}{(z \cdot z\rq{})^{n + l}} \ .
\ee
Using \eqref{PrimaryCondIcheck} it is straightforward to obtain the recurrence relation 
\begin{align}
N_{\I, n} =-2 n(\D+n+l-1) N_{\I, n-1} \label{NIiteration} \ .
\end{align}
Iterating equation (\ref{NIiteration}) up to  $N_{\I, 0}=1$, we obtain 
\be
N_{\I,n}=  (-2)^n n! (\D+l)_n\ .
\label{norm_I}
\ee

\subsection{Type II}
We now wish to demonstrate that a descendant of type \II becomes a primary when $\D=\D^\star_{\II, n}$,
\be
 (K\cdot z\rq{}) (D_z \cdot P)^n  | \D, l \,;z \rangle= 0\ , \qquad\qquad \textrm{if }\D= \D^{\star}_{\II,n}\equiv  l+d-1-n\ . \label{PrimaryCondII}
\ee
Using \eqref{commutatorKPzn} and replecing $z$ with $D_z$, we obtain the following commutator,
\begin{align}
[ (K\cdot z\rq{}), (P \cdot D_z)^n ] =2  i\, n\, (P \cdot D_z)^{n-1}  [(z\rq{} \cdot D_z) (D+ i (n-1)) - (z\rq{} \cdot J \cdot D_z)] \ . \label{commutatorKPDzn}
\end{align}
The eigenvalue of $(z\rq{} \cdot J \cdot D_z)$ on a primary state is
\be
 (z\rq{} \cdot J \cdot D_z)| \D, l \,;z \rangle=i (l - 2 + d)  (z\rq{} \cdot D_z)| \D, l \,;z \rangle \ .
\ee
Therefore when we act with \eqref{commutatorKPDzn} on $| \D, l \,;z \rangle$ we find
\be \label{PrimaryCondIIcheck}
(K\cdot z\rq{}) (D_z \cdot P)^n| \D, l \,;z \rangle =-2 n  (\D- l  - d+ 1+ n)  (z\rq{} \cdot D_z)(P \cdot D_z)^{n-1} | \D, l \,;z \rangle \ ,
\ee
which proves \eqref{PrimaryCondII}.

The norm $N_{\II, n}$ can be defined as
\be
 N_{\II, n} \equiv \frac{\langle \D +n , l+n; z\rq{} | \D+n, l+n; z \rangle}{(z \cdot z\rq{})^{l-n}}=\frac{\langle \D , l ;z\rq{} |\frac{(K\cdot \overset{_\leftarrow}{D}_{z\rq{}})^n}{(2-h-l)_n(-l)_n} \frac{(P \cdot \overset{_\rightarrow}{D}_z)^n}{(2-h-l)_n(-l)_n}| \D, l;z \rangle}{(z \cdot z\rq{})^{l-n}} \ .
\ee
where $\overset{_\leftarrow}{D}_z$  and $\overset{_\rightarrow}{D}_z$ act respectively on the left and on the right.  Using the following identity
\be
\frac{(D_z\cdot D_{z\rq{}})^n}{(2-h-l)_n^2 (-l)_n^2} (z \cdot z\rq{})^{l} =\frac{(3-d-l)_n}{(-l)_n } \frac{ 
   (d+ 2l-2)}{(d+ 2l- 2n-2) } (z \cdot z\rq{})^{l-n} \ ,
\ee
and applying recursively \eqref{PrimaryCondIIcheck} (with $z\rq{} \rightarrow \overset{_\leftarrow}{D}_{z\rq{}}$) as we did for the norm of type \I, we obtain 
\be
N_{\II,n} = 
 (-2)^n n!(\D-d-l+2)_n \frac{(3-d-l)_n}{(-l)_n
    } 
   \frac{ 
   (d+ 2l-2)}{(d+ 2l- 2n-2) } \ .
\label{norm_II}
\ee
\subsection{Type III}
The descendants of type \III are more complicated than the ones of the previous two cases.
In fact at level $2n$,  there are $1+\min(l,n)$ multiplets of spin $l$, which can be written as
\be \label{stateIIIj}
(P^2)^{n-j} (P\cdot z)^j (P\cdot D_z)^j | \D, l \,;z \rangle \ ,
\ee
where $j=0,1,\dots,\min(l,n)$. To find which is the correct linear combination of the states (\ref{stateIIIj}), we impose 
\be
 (K\cdot z\rq{}) |\D+2n,l\, ;z\rangle =0\ , \qquad\qquad \textrm{if }\D= \D^{\star}_{\III,n}\equiv h-n\ . \label{PrimaryCondIII}
\ee
From numerical experiments, we conjecture the following form
\be
|\D+2n,l \, ;z\rangle =\Dcal_{\III,n}|\D,l\, ;z \rangle \equiv  \sum_{j=0}^{\min(l,n)} a(j) \,(P^2)^{n-j} (P\cdot z)^j (P\cdot D_z)^j 
|\D,l \, ;z \rangle \label{stateIII}
\ee
with coefficients
\be
a(j)\equiv 
\frac{l! n!}{j! (l-j)! (n-j)!}
\frac{ (-2)^j }{(h+l+n-j-1)_j}
 \frac{1}{(2-h-l)_j (-l)_j} \ .
\label{aj}
\ee
Formula  \eqref{stateIII} can be also written as \eqref{stateIIIother}, which is given in the main text.

We would like to compute the norm $N_{\III,n}$ defined by
\be
\langle \D+2n,l\,;z\rq{} |\D+2n,l\, ;z\rangle=
N_{\III,n} (z \cdot z\rq{})^l \ .
\ee
We did not find a closed formula.  However when $l=0$, the only multiplet is $(P^2)^{n}|\D,0  \rangle$, therefore we can compute the norm 
\be \label{normIIIl=0}
N_{\III, n} \big|_{l=0} =16^n n! (h)_n  (\Delta )_n (\Delta  -h+1)_n \ .
\ee
Moreover by inspection we found, for example,
{\scriptsize
\be
\!\!\!\!\!
\begin{array}{|c | l |}
\hline
\phantom{ \Bigg{(}}n\phantom{ \Bigg{(}}& \; N_{\III, n} \\ 
\hline
\phantom{ \Bigg{(}}1\phantom{ \Bigg{(}} &
\dfrac{16 (h-1) (\Delta -h+1) (h+l)}{(h+l-2) (h+l-1)^2}
   \left[(\Delta -h+1) \left((h+l-1)^2-(h-1)\right)+(h-2) (h+l-1)^2 \right] \  , \\
\hline
\phantom{ \Bigg{(}}2\phantom{ \Bigg{(}}&
\dfrac{512 (h-1) h (\Delta -h+2) (\Delta -h+1) (h+l+1) }{(h+l-3) (h+l-2) (h+l-1)^2 (h+l)}
\big[(\Delta -1) \Delta  (h+l-2) (h+l-1)^2 (h+l) \qquad \nonumber\\
\phantom{ \Bigg{(}} &+2 h (\Delta -h+2) \left((h-1) (\Delta -h+1)-(\Delta -1) (h+l-1)^2\right) \big] \ , \\\hline
\phantom{ \Bigg{(}}3\phantom{ \Bigg{(}}&
-\dfrac{24576 (h-1) h (h+1) (-\Delta +h-3) (-\Delta +h-2) (-\Delta +h-1) (h+l+2)}{(h+l-4) (h+l-3) (h+l-2) (h+l-1)^2 (h+l) (h+l+1)}\Big[\Delta  \left(\Delta ^2-1\right) l^6+2 (\Delta -1) \times
\nonumber \\
\phantom{ \Bigg{(}} &\times (h-1) l \left(-6 \Delta  (\Delta +2)+3 (\Delta +1) (\Delta +2) h^4-6 (\Delta +2) (3 \Delta +2) h^3+(5 \Delta  (4 \Delta +7)+6) h^2+(\Delta  (23 \Delta +68)+36) h\right)
\nonumber \\
\phantom{ \Bigg{(}} &
+4 (\Delta -1) \Delta  (h-1) l^3 (h (-13 \Delta +(5 \Delta +8) h-16)-3 (\Delta +3))+\Delta  (\Delta +1) (\Delta +2) (h-4) (h-3) (h-2) (h-1) h (h+1)
\nonumber \\
\phantom{ \Bigg{(}} &
+(\Delta -1) l^2 \left(\Delta ^2 (3 h (h (h (5 h-26)+28)+9)-26)
+\Delta  (3 h ((h-1) h (11 h-37)+36)-56)+6 (h-3) (h-2) h (h+1)\right) 
\nonumber \\
\phantom{ \Bigg{(}} &
+6 \Delta  \left(\Delta ^2-1\right) (h-1) l^5+(\Delta -1) \Delta  l^4 (7 \Delta +3 h (-11 \Delta +(5 \Delta +6) h-12)+1)\Big]\ , \\
\hline
\end{array}
\ee
}
Motivated by these results, we came to the conjectured result
\begin{align}
 \frac{1}{N_{\III,n} }
   &\approx -\frac{n}{16^n (n!)^2 (h)_n(1-h)_n}
   \frac{(h+l-n-1)(h+n-1)}{(h+l+n-1)(h-n-1)} 
   \; \frac{1}{
\Delta -h+n} \ .
\label{norm_III}
\end{align} 
It would be nice to prove this result analytically.
\section{Computation of \texorpdfstring{$M_A$}{MA}}
\label{App:RA}
In this section we explain one possible way to compute the coefficients $M_{A}$ defined in \eqref{DA-MA} for the scalar-scalar OPE, and in \eqref{DAt=MAt} for the vector-scalar OPE.
\subsection{Scalar-scalar OPE}
\label{App:RAscalarconformalblocks}
We first compute $M_{A}$ defined in \eqref{DA-MA} by
\be
\Dcal_A \frac{(-x\cdot z)^l}
{(x^2)^{\a}} = M_A   \frac{(-x\cdot z)^{ l_A}}
{(x^2)^{\a_A}} \ ,
\ee
where $\a=\frac{\D+\D_{12}+l}{2}$ and $\a_A=\frac{\D+n_A+\D_{12}+ l_A}{2}$. Throughout this appendix we will adopt the notation
\be
A=T,n \ , \qquad
\left\{
\begin{array}{ll}
T&=\I,\II,\III \\
n&=1,2,\dots
\end{array}
\right. \ .
\ee
The method we use to compute $M_{A}$ relies on the observation that the differential operators $\Dcal_{T,n}$ can be expressed as $(\Dcal_{T,1})^n$ (this statement is exact only for $T=\I,\II$, while for $T=\III$ one has to slightly correct it). Hence roughly speaking, it is sufficient to act with the differential operator $\Dcal_{T,1}$ on $\frac{(-x\cdot z)^l}
{(x^2)^{\a}}$ only once, and then \lq\lq{}multiply\rq\rq{} $n$ times the resulting coefficient. 

The exact procedure is as follows.
We define a coefficient $m_T(j)$ for the types $T=\I,\II,\III$  by
\begin{equation}
\begin{array}{clll}
\Dcal_{\I,1}&\dfrac{(-x\cdot z)^{l+j}}{(x^2)^{\a+j}} & =m_{\I}(j) &\dfrac{(-x\cdot z)^{l+j+1}}{(x^2)^{\a+j+1}} \ ,\\
\Dcal_{\II,1}&\dfrac{\; (-x\cdot z)^{l-j}}{(x^2)^{\a}}& =m_{\II}(j) &\dfrac{(-x\cdot z)^{l-j-1}}{(x^2)^{\a}} \ ,\\
\Vcal_{j} &\dfrac{(-x\cdot z)^{l}}{(x^2)^{\a + j}}& =m_\III(j)& \dfrac{(-x\cdot z)^l}{(x^2)^{\a+j+1}} \ ,
\end{array}
\end{equation}
where we opportunely shifted $\a$ and $l$ in such a way to obtain $M_{A}$ simply as the product
\be \label{recurrenceMAscalar}
M_{T,n}=\prod_{j=0}^{n-1} m_T(j)  \ .
\ee
We find the coefficients
\begin{equation}
\begin{array}{cl}
m_{\I}(j)&=2 i (\alpha +j)  \ , \phantom{\dfrac{a}{b}} \\
m_{\II}(j)&=-\dfrac{i (2 h-j+l-3) (-\alpha +h-j+l-1)}{h-j+l-2} \ ,\\
m_{\III}(j)&=\dfrac{4 (h-j-2) (h+j-1) (\alpha +j) (-\alpha +h-j+l-1)}{(h-j+l-2) (h+j+l-1)} \ ,\\
\end{array}
\end{equation}
which, using (\ref{recurrenceMAscalar}), give \eqref{Mscalar}.
\subsection{Scalar-vector OPE}
\label{App:The residue RA for vector conformal blocks}
We now generalize the previous procedure to compute the matrix $M_A$ defined in \eqref{DAt=MAt}.
We first compute the $2 \times 2$ matrix $m_T(j)$ for the types $T=\I,\II,\III$ 
\begin{align}
\Dcal_{\I,1}\;\frac{t^{(q)}_{l+j}(x,z, z_1)}{(x^2)^{\a+j}} & =\sum_{q\rq{}=1}^2 \left(m_{\I}(j)\right)_{q q\rq{}} \frac{t_{l+j+1}^{(q\rq{})}(x,z, z_1)}{(x^2)^{\a+j+1}} \ ,\\
\Dcal_{\II,1}\;\frac{t^{(q)}_{l-j}(x,z, z_1)}{(x^2)^{\a}}& =\sum_{q\rq{}=1}^2 \left(m_{\II}(j)\right)_{q q\rq{}} \frac{t_{l-j-1}^{(q\rq{})}(x,z, z_1)}{(x^2)^{\a}} \ ,\\
\Vcal_{j} \; \;\;\frac{t^{(q)}_l(x,z, z_1)}{(x^2)^{\a + j}}& =\sum_{q\rq{}=1}^2 \left(m_\III(j) \right)_{q q\rq{}} \frac{t_{l}^{(q\rq{})}(x,z, z_1)}{(x^2)^{\a+j+1}} \ .
\end{align}
The resulting matrices $m_{T}(j)$ are 
\begin{equation} \label{mTexplicit}
\begin{array}{cl}
m_{\I}(j)&=-2i \left(
\begin{array}{cc}
   j+\alpha -1 & -1/2 \\
 0 &   j+\alpha -2 \\
\end{array}
\right) \ ,\\
m_{\II}(j)&=\frac{i}{h-j+l-1}\left(
\begin{array}{cc}
 -h+\alpha +p_j & \frac{j-l}{2} \\
 \frac{2 (h-1) (\alpha -1)}{j-l-1} & \frac{(j-l) p_j}{j-l-1} \\
\end{array}
\right) \ ,\\
m_{\III}(j)&=\frac{4 (\alpha +j-1) (\alpha +j)}{(h-j+l-2) (h+j+l-1)}\left(
\begin{array}{cc}
 \frac{q_j}{j+\alpha -1} & \frac{l \left(h^2+(-2 j+l-2 \alpha -1) h+j (j+3)-l+2 \alpha \right)}{2 (j+\alpha -1) (j+\alpha )} \\
 2 (h-1) & \frac{h^2-(2 j+2 \alpha +1) h+j (j+3)+2 \alpha +q_j}{j+\alpha } \\
\end{array}
\right)\ ,\\
\end{array}
\end{equation}
where
\be
\begin{array}{lcl}
p_j&\equiv&\alpha +2 h^2+h (-2 \alpha -3 j+3 l)+j^2+j (\alpha -2 l+1)+l^2-\alpha  l-l-1 \ , \\
q_j&\equiv&-2 \alpha +h^3-h^2 (\alpha +j-l+3)+h (3 \alpha -(j-2) j-2 l+2)-2 j+l \\
 &&+j (j+1) (\alpha +j-l) \ .\\
\end{array}
\ee
The matrix $M_{T,n}$ can be obtained by multiplying $n$ matrices $m_T(j)$ as follows
\be \label{recurrenceMA}
M_{T,n}=m_T(0) m_T(1) \cdots m_T(n-1)  \ .
\ee
For the type \I, one can diagonalize $m_{\I}(j)$ using a matrix that does not depend on $j$, therefore it is trivial to find a closed form for $M_{\I,n}$. In the other cases this procedure does not work. However it is not hard to guess a closed form for $M_{A}$ and then use  formula (\ref{recurrenceMA}) to prove the guess by induction. The final result is
\begin{align}
\begin{split}
 \label{MAfinalresult}
M_{\I,n}&=(-2i )^n (\alpha )_{n-1}\left(
\begin{array}{cc}
 n+\alpha -1 & -\frac{n}{2} \\
 0 & \alpha -1 \\
\end{array}
\right) \ , 
\\
M_{\II,n}&=\tfrac{(-i)^n (-2 h-l+3)_{n-1} (-h-l+\alpha +1)_{n-1}}{(-h-l+2)_n}\left(
\begin{array}{cc}
 P_n & \tfrac{n}{2} (n-l)\\
 \frac{2n (1-\alpha) (h-1)  }{l}  & \frac{(l-n)(P_n-n \alpha+n h)}{l} \\
\end{array}
\right) \ ,
\\
M_{\III,n}&=\tfrac{ (h-n)_{2 n-1} (\alpha )_{n-1} (h+l-n-\alpha +1)_{n-1}}{2^{-2 n}(h+l-n-1)_{2 n}} \left(
\begin{array}{cc}
  (\alpha +n-1) Q_n &   l/2 \; (R_n+n l)\\
2 n (\alpha -1) (\alpha +n-1)&  (\alpha -1) (Q_n+R_n) \\
\end{array}
\right)  \ ,
\end{split}
\end{align}
where 
\be
\begin{array}{ll}
P_n&\equiv 2 h^2 + l^2 + n + (2 + n) \a - l (2 + n + \a) + 
 h (3 l - 2 (1 + n + \a)) \ , \\
Q_n&\equiv \alpha +h (h+l-n-1)+\alpha  (n-h)-l  \ , \\
R_n&\equiv n (1 + h  - n - 2 \a)   \ .
\end{array}
\ee

\section{Conformal Blocks at Large \texorpdfstring{$\D$}{Delta}}
\label{App:CBatLargeDelta}

We want to have a general set up to study conformal blocks at large $\D$ in the case of external and exchanged operators with generic spin and belonging to the symmetric and traceless representation.
\subsection{Setup for Conformal Blocks with Spin}
\label{CBwithspin}
Given three traceless and symmetric primary operators $\Ocal, \Ocal_1, \Ocal_2$ with conformal dimensions $\D, \D_1, \D_2$ and spins $l, l_1,l_2$, we can define the following leading OPE 
 \be \label{OPEwithSPIN}
\Ocal(x,z)\Ocal_1(0,z_1)\sim \frac{\Ocal_2(0,\partial_{z_2})}{(x^2)^{\a}} \sum_{q}
c_{12\Ocal}^{(q)} \; t^{(q)}_l( x,z,z_1,z_2)  \ ,
\ee
where $\a\equiv \frac{\D+\D_1-\D_2+l+l_1+l_2}{2}$ and  $c_{12\Ocal}^{(q)}$ are the OPE coefficients. The tensor structures   $ t^{(q)}_l(x,z,z_2,z_3)$  must be  Lorentz invariant and satisfy 
\be
t^{(q)}_l(\m x, \l z, \l_1 z_1,\l_2 z_2)= \m^{l+l_1+l_2} \l^l \l_1^{l_1}\l_2^{l_2} \; t^{(q)}_l( x,z,z_1,z_2) \  .
\ee
It is also possible to reconstruct the full three point function from the leading OPE \cite{Mack:1976pa,OsbornCFTgeneraldim,SpinningCC}
\begin{align} \label{3ptFunctionSpin}
 &  \langle \Ocal(x_0,z)\Ocal_1(x_1,z_1)\Ocal_2(x_2,z_2) \rangle \nonumber\\
 &  \qquad  = \frac{ \sum_{q} c_{12\Ocal}^{(q)} \; t^{(q)}_l\left(\tilde x_{01} , I(x_{02})\cdot z,I(x_{12}) \cdot z_1,z_2 \right)}{(x_{02}^2)^{\frac{\D+\D_2-\D_1+l+l_1+l_2}{2}}  (x_{12}^2)^{\frac{\D_1+\D_2-\D+l+l_1+l_2}{2}} (x_{01}^2)^{\frac{\D+\D_1-\D_2+l+l_1+l_2}{2}}} \ ,
\end{align}
where $\tilde x_{01} \equiv x_{02}x_{12}^2-x_{12}x_{02}^2$. Notice that in (\ref{3ptFunctionSpin}) the variables $z_1$ and $z_2$  appear in a non symmetric way. This is not surprising since the  structures $ t^{(q)}_l$ are defined from the OPE 
\eqref{OPEwithSPIN} which treats differently the   operators $\Ocal_1$ and $\Ocal_2$. From (\ref{3ptFunctionSpin}) one can also write the leading OPE in the other channels. They can be all defined using the structures $t^{(q)}_l$, once they are transformed in the right way, for example 
\be \label{OPEO1O2}
\Ocal_1(x,z_1)  \Ocal_2(0,z_2) \sim  \frac{1}{ l!\, (h-1)_l  } \frac{\Ocal(0,D_z)}{(x^2)^{\frac{\D_1+\D_2-\D+l+l_1+l_2}{2}}} \sum_{q} c_{12\Ocal}^{(q)} \; t^{(q)}_l(-x,z,I(x)z_1,z_2) \ .
\ee

Let us now consider  the conformal block decomposition of a four point function of symmetric traceless tensor operators $\mathcal{F}_4(\{x_i,z_i\})=
\langle\Ocal_1(x_1,z_1)\Ocal_2(x_2,z_2)\Ocal_3(x_3,z_3)\Ocal_4(x_4,z_4) \rangle$, where each operator $\Ocal_i$ has spin $l_i$.
We can write 
\begin{align}
\label{CBwithSPIN1}
\mathcal{F}_4(\{x_i,z_i\})&=\sum_{\Ocal} \sum_{p,q} c_{12\Ocal}^{(p)} c_{34\Ocal}^{(q)} G_{\D l}^{(p, q)}(\{x_i,z_i\}) +\dots \\
&={\sum_{\Ocal} \sum_{p,q}} %
\raisebox{1.7em}{$\xymatrix@=6pt{{\Ocal_1}\ar@{-}[rd]& & & &\Ocal_3 \ar@{-}[ld]   \\  
& *+[o][F]{\mbox{\tiny p}}  \ar@{=}[rr]^{\displaystyle \Ocal } & & *+[o][F]{\mbox{\tiny q}}  &  \\
\Ocal_2 \ar@{-}[ru]& & & &\Ocal_4 \ar@{-}[lu]}$}+\dots\ ,  
\end{align}
where the $\dots$ stand for the contribution of  other  irreducible representations of $SO(d)$ that are not symmetric traceless tensors and that can be exchanged in this correlator. We want to study the leading behavior in $\D$ of $G_{\D l}^{(p, q)}(\{x_i,z_i\})$. 

\subsection{Conformal Blocks in the Embedding Formalism}
\label{Four Point Function of Vector Operators}
To simplify the notation in this appendix we introduce the embedding formalism. We will uplift each $x\in \mathbb{R}^d$ to a $P \in \mathbb{M}^{d+2}$ such that $P^2=0$. To find  $x \in \mathbb{R}^d$ given $P$ we will have just to consider a projection of $P$ onto the Poincar\'e section, in light cone coordinates we can write $P_x=(1,x^2,x^\m)$ with $\m=1,\dots,d$. To be consistent we will need to use also new auxiliary vectors $Z_{z,x}=(0, 2 x\cdot z, z^{\mu})$ defined on $\mathbb{M}^{d+2}$ to encode the tensor structures. The main reason why we want to work in the embedding is that the action of the conformal group (in $d$ dimension) is linear on $\mathbb{M}^{d+2}$, so that any scalar object in the embedding space is also a conformal invariant and vice-versa. More details about it can be found in \cite{SpinningCC} and references therein.

As in the scalar case, we can use conformal symmetry to fix a generic conformal block $G_{\D l}^{(p,q)}(P_i,Z_i)$ %
 up to functions of two real variables. When the external operators are tensor we have many of such functions, each one of them multiplied by a different tensor structure
 \cite{SpinningCC}
\be 
G_{\D l}^{(p,q)}(P_i,Z_i)=\frac{ \left(\frac{P_{24}}{P_{14}} \right)^{\frac{\D_1-\D_2}{2}}  \left(\frac{P_{14}}{P_{13}} \right)^{\frac{\D_3-\D_4}{2}}}{(P_{12})^{\frac{\D_1+\D_2}{2}}(P_{34})^{\frac{\D_3+\D_4}{2}}} \sum_{s} g^{(p,q)}_{\D,l,s}(r, \eta) \mathcal{Q}^{(s)}(\{P_i,Z_i\}) \ , \label{CB:structuresApp}
\ee
where $P_{i j}=-2 P_i \cdot P_j$  and $\mathcal{Q}^{(s)}$ are all the possible scalar structures constructed with $P_i$ and $Z_i$ with the following property 
\be
\mathcal{Q}^{(s)}(\{\l_i P_i,\a_i Z_i+\b_i P_i\})=\mathcal{Q}^{(s)}(\{P_i,Z_i\}) \prod_{i} (\a_i)^{l_i},
\label{Q_homogeneous}
\ee
where $l_i$ is the spin of the operator $\Ocal_i$ in the four point function.
A convenient way to generate $\mathcal{Q}^{(s)}$ is by using the building blocks $V_{i,jk}$ and $H_{i j}$ of  \cite{SpinningCC}, each of which 
is invariant under the transformation $P_i\to \lambda_i P_i, Z_I\to Z_i+\beta_i P_i$: 
\begin{align}
V_{i,jk}&=\frac{(Z_i\cdot P_j)(P_i\cdot P_k)-(Z_i\cdot P_k)(P_i\cdot P_j)}{\sqrt{-2(P_i\cdot P_j)(P_j\cdot P_k)(P_k\cdot P_i)}}  \ , \\
H_{i j}&=\frac{(Z_i\cdot Z_j)(P_i\cdot P_j)-(Z_i\cdot P_j)(P_i\cdot Z_j)}{(P_i\cdot P_j)} \ .
\end{align}
Note that we have $V_{i,jk}=-V_{i,kj}$ and $H_{ij}=H_{ji}$.
Since we are interested in a four point function we also have the constraint 
\be
(P_2\cdot P_3)(P_1\cdot P_4)V_{1,23}+(P_2\cdot P_4)(P_1\cdot P_3)V_{1,42}+(P_3\cdot P_4)(P_1\cdot P_2)V_{1,34}=0  \ ,
\ee 
which means that for each point $i$ we only have two independent choices of $V_{i,jk}$. 
Projecting all  the scalar combinations into the Poincar\'e section ($P\to P_x, Z\to Z_{z,x}$)
\be
P_i \cdot P_j \to -\frac{1}{2}  x^2_{i j} \ , \qquad Z_i \cdot P_j \to -z_i \cdot x_{i j} \ ,  \qquad  Z_i\cdot Z_j \to z_i \cdot z_j \ ,
\ee
the structures can be written in terms of spacetime coordinates after the replacement
\begin{align} \label{HVpoincaresection}
H_{ij}& \to \textfrak{h}_{ij}\equiv (z_i \cdot z_j)   -2 \frac{(z_i \cdot x_{i j}) (z_j \cdot x_{i j})}{x^2_{i j}}=z_i I(x_{i j}) z_j \ ,\\
V_{i, jk}&\to \textfrak{v}_{i, jk}\equiv(z_i \cdot \hat x_{i j} )\frac{|x_{i k}|}{|x_{j k}|}-(z_i \cdot \hat  x_{i k} )  \frac{ |x_{i j}|}{|x_{j k}|} \ .
\end{align}
In the case of three scalars and one vector, the resulting basis for independent structures is
\be
\label{basisQs}
\left\{
\begin{array}{l}
\mathcal{Q}^{(1)}(\{P_i\},Z_1)=V_{1,23}\\
\mathcal{Q}^{(2)}(\{P_i\},Z_1)=V_{1,34}
\end{array}
\right. \ .
\ee
The basis (\ref{basisQs}) reduces to $Q^{(s)}(\{x_i\},z_1)$ in (\ref{basisQs1}) in the main text once we use (\ref{HVpoincaresection}).
\subsection{Casimir Equation at Large \texorpdfstring{$\D$}{Delta}}
\label{App:CasimirEquation}
Another reason to use the embedding formalism is that it simplifies the computations for the Casimir equation. In fact in the embedding space the generators of conformal transformations are simple rotations
\be
(J_{(P,Z)})_{AB}\equiv P_A \partial_{P_B}-P_B \partial_{P_A}+Z_A \partial_{Z_B}-Z_B \partial_{Z_A} \ ,
\ee
and therefore the Casimir can be simply written as
\be
\mathcal{C}\equiv \frac{1}{2}(J_{(P_1,Z_1)}+J_{(P_2,Z_2)})_{AB}(J_{(P_1,Z_1)}+J_{(P_2,Z_2)})^{AB} \ .
\ee
The Casimir equation can be formulated as 
\be
\mathcal{C} \; G_{\D l}^{(p , q)}(\{X_i,Z_i\})= c_{\D l} G_{\D l}^{(p, q)}(\{X_i,Z_i\})  \ ,\label{casimir eq gen}
\ee
where $G_{\D l}^{(p, q)}(\{X_i,Z_i\})$ has to be replaced by (\ref{CB:structuresApp}). To get the leading behavior in $\D$ of (\ref{casimir eq gen}) we replace  $g^{(q)}_{\D,l,s}(r, \eta)=(4r)^{\D} h^{(q)}_{\D,l,s}(r, \eta)$. It is easy to see that the terms proportional to $\D^2$ exactly cancel from the two sides of the equation, and these terms are the ones generated by a second derivative in $r$. The equation is then just linear in $\D$ and the highest degree term gives rise to a first order differential equation of the form
\be
\partial_r F_s ^{\phantom{s}t}(r,\eta)= \sum_{s\rq{}} \mathcal{M}_{s\, s\rq{}}(r,\eta) F_{s\rq{}}^{\phantom{s}t}(r,\eta) \ , 
\ee
where the sum over $s\rq{}$ runs over the possible tensor structures of the four point function.

\subsection{OPE Limit of Conformal Blocks}
\label{Conformal_Block_at_large_Delta}

To find the functions $f^{(p,q)}_{l, s\rq{}}(\eta) \equiv h^{(p,q)}_{\infty,l,s}(0,\eta)$ we need to study the leading behavior of $G_{\D l}^{(p,q)}$ for $r \rightarrow 0$. 
From the three point function (\ref{3ptFunctionSpin}) we can read off the leading order (in $x_{12}$) OPE between two vector operators, 
\begin{align} \label{OPEwithSPIN1}
&\Ocal_1(x_1,z_1)\Ocal_2(x_2,z_2)  \nonumber\\
& \qquad \approx \frac{1}{l! (h-1)_l} \frac{\Ocal(x_{2},D_z)}{(x_{12}^2)^{\frac{\D_1+\D_2-\D+l+l_1+l_2}{2}}}\sum_{q} c_{12\Ocal}^{(q)}  t_{l}^{(q)}(-x_{12}, z,I(x_{12})\cdot z_1,z_2) \ ,
\end{align}
where $\Ocal$ has spin $l$ and conformal dimension $\D$.
Now we consider the four point function $\mathcal{F}_4(\{x_i,z_i\})$, we take the leading OPE (\ref{OPEwithSPIN1}) in the channels $1-2$ and $3-4$ and then we write the remaining two point function as in (\ref{def:2ptspin}), to obtain the following result
\begin{align} \label{formula:CBsmallr1}
G_{\D l}^{(p,q)}(\{x_i,z_i\}) 
&\approx \frac{(D_z \cdot I(x_{24}) \cdot D_{z\rq{}})^l }{ [l! (h-1)_l ]^2}   \frac{t^{(p)}_l (x_{12},z \rq{},I(x_{12}) \cdot z_1, z_2)t^{(q)}_l(x_{34},z ,I(x_{34}) \cdot z_3,z_4)}{ (x^2_{12})^{\frac{\D_1+\D_2-\D+l_1+l_2+l}{2}}(x^2_{34})^{\frac{\D_3+\D_4-\D+l_3+l_4+l}{2}}(x^2_{24})^\D } \ .
\end{align}
In the limit of $x_{12},x_{34}\rightarrow 0$ we obtain that to the leading order $x_{13}\approx x_{23} \approx x_{2 4} \approx x_{14}$ and moreover we have $(4 r)^2 \approx \frac{x_{12}^2 x_{34}^2}{(x_{24}^2)^2} $. Using this simplifications and applying the Todorov operator $ D_{z\rq{}}$ on the left structure of (\ref{formula:CBsmallr1}) we obtain the following formula 
\begin{align}
G_{\D l}^{(p,q)}(\{x_i,z_i\}) 
\approx (4 r)^{\D} \frac{ t^{(p)}_l( \hat x_{12},I(x_{24})\cdot D_z,I(x_{12}) \cdot  z_1, z_2)t^{(q)}_l(\hat x_{34},z,I(x_{34}) \cdot z_3,z_4)}{l! (h-1)_l \qquad (x^2_{12})^{\frac{\D_1+\D_2}{2}} \qquad (x^2_{34})^{\frac{\D_3+\D_4}{2}}  }  \ . \label{smallrCBspin}
\end{align}
Taking the OPE limit of (\ref{CB:structuresPhysical}) and replacing $g^{(p,q)}_{\D,l,s}(r, \eta)=(4r)^\D h^{(p,q)}_{\D,l,s}(r, \eta)$ we find
\be 
G_{\D l}^{(p,q)}(\{x_i,z_i\})\approx \frac{(4r)^\D}{(x_{12})^{\frac{\D_1+\D_2}{2}}(x_{34})^{\frac{\D_3+\D_4}{2}}} \sum_{s} h^{(p,q)}_{\D,l,s}(r, \eta) Q^{(s)}(\{x_i,z_i\}) \ . \label{CB:structuresrsmall}
\ee
To completely fix the functions $f^{(p,q)}_{l,s}(\eta)$ we need to compare formula (\ref{smallrCBspin}) with (\ref{CB:structuresrsmall}) for $r$ approaching zero, 
\begin{align}
\!\!\!\! \frac{ t^{(p)}_l( \hat x_{12}, 
   I(x_{24})\!\cdot \! D_z ,I(x_{12}) \cdot  z_1, z_2)t^{(q)}_l(\hat x_{34},z,I(x_{34})\! \cdot \!z_3,z_4) 
}{l! (h-1)_l}
 \approx  \sum_{s} f^{(p,q)}_{l,s}(\eta) Q^{(s)}(\{x_i,z_i\}) \ .\label{smallr:eq.forf1}
\end{align}

The aim of what follows is to clarify how we get a matching of the two sides of the formula (\ref{smallr:eq.forf1}). To do so we first study how the structures $Q^{(s)}(\{x_i,z_i\})$ in the right hand side of (\ref{smallr:eq.forf1}) behave for small $r$. We recall that $Q^{(s)}(\{x_i,z_i\})$ can be always expressed in terms of the two building blocks $\textfrak{v}_{i,jk}$ and $\textfrak{h}_{i j}$. This means that it is sufficient to study their behavior at small $r$, to understand how the whole set of possible structures behave. We obtain, for small $r$, 
\be
\begin{array}{l l  l}
\textfrak{v}_{a,b i}\approx z_a \cdot \hat x_{a b} \ , 
& \textfrak{v}_{i,j a}\approx z_i \cdot \hat x_{i j} \ ,&\\
\textfrak{v}_{i,1 2}\approx - z_i \cdot I(x_{2 4})\cdot \hat x_{12} \ , &
\textfrak{v}_{a,3 4}\approx -z_a \cdot I(x_{2 4}) \cdot \hat x_{34} \ ,&\\
\textfrak{h}_{12}\approx z_1 \cdot I(x_{12}) \cdot z_2  \ ,\qquad \qquad 
&\textfrak{h}_{34}\approx z_3\cdot I(x_{34}) \cdot z_4  \ , \\
\textfrak{h}_{a i}\approx z_a \cdot I(x_{2 4}) \cdot z_i  \ ,
\end{array} \label{smallr:VH}
\ee
where $a,b=1,2$ and $i,j=3,4$.

We now want to understand how it is possible to find the structures listed above on the left hand side of (\ref{smallr:eq.forf}). 
The simplest building blocks that appear in  the left hand side of (\ref{smallr:eq.forf}) are all the scalar combination of left variables $\hat x_{12},I(x_{12}) \cdot z_1, z_2$ only, and that of the the right variables  $\hat x_{34},I(x_{34}) \cdot  z_3, z_4$ only. These combinations generates $\textfrak{v}_{a,b i},\textfrak{v}_{i,j a},\textfrak{h}_{12},\textfrak{h}_{34}$. We can also build more complicates structures that mix the left variables with the right ones. In fact when a combination $ \hat x_{12}\cdot  I(x_{24}) \cdot D_z,z_1\cdot I(x_{12}) \cdot I(x_{24}) \cdot D_z,z_2\cdot I(x_{24}) \cdot D_z$ appears on the left structure, it then acts on one of the three possible combinations on the right $\hat x_{34}\cdot z,z_3\cdot I(x_{34}) \cdot z,z_4\cdot z$. This action creates  structures that can be written as linear combinations of (\ref{smallr:VH}), with the single exception of $\hat x_{12}I(x_{24}) \hat x_{34}$. 
In fact,
\be
\eta \approx - \hat x_{12}I(x_{24}) \hat x_{34} \ ,
\ee
thus this combination actually gives the dependence on $\eta$, which is needed for the identifications of the functions $f^{(p,q)}_{l,s}(\eta)$.

Notice that in doing the matching  (\ref{smallr:eq.forf}) we need some formula of the kind (\ref{formula:gegenb-projector}). Actually in a generic case we will need to compute some expression of the form
\be \label{general_gegenb}
 \Fcal(\{u_i, v_i\},x,y)=\frac{(u_1 \cdot D_z)\cdots (u_j \cdot D_z)(x \cdot D_z)^{l-j}}{l! (h-1)_l} (y \cdot z)^{ l-k} (v_1 \cdot z)\cdots  (v_k \cdot z) \ .
\ee
The main idea to compute (\ref{general_gegenb}) is that one can always rewrite it as a set of derivative acting on $\frac{(x \cdot D_z)^{l}}{l! (h-1)_l} (y \cdot z)^{ l}$ and then apply \eqref{formula:gegenb-projector} as follows
\begin{align}
 \Fcal(\{u_i, v_i\},x,y) &=\left( \prod_{i=1}^j \frac{(u_i \cdot \partial_{x})}{l-i+1} \right)\left( \prod_{i=1}^k \frac{(v_i \cdot \partial_{y})}{l-i+1} \right) \frac{(x \cdot D_z)^{l}(y \cdot z)^{ l}}{l! (h-1)_l}  \nonumber \\
&=\left( \prod_{i=1}^j \frac{(u_i \cdot \partial_{x})}{l-i+1} \right)\left( \prod_{i=1}^k \frac{(v_i \cdot \partial_{y})}{l-i+1} \right)  \frac{l!(x^2 y^2)^{l/2} C_l^{h-1}(\hat x \cdot \hat y)}{2^l\, (h-1)_l }  \ .
\end{align}
Moreover one can use the identities for derivatives of Gegenbauer polynomial such as
\be
\partial_x C_l^{h}(x)= 2 h \; C_{l-1}^{h+1}(x) \ .
\ee
Putting together all these ingredients one can find the functions $f^{(p,q)}_{l,s}(\eta)$ that define the OPE limit of the conformal blocks.

\section{Definitions for the CB with One External Conserved Current}
\label{App.ConsCurr}
In this appendix we write down the formulas for $\tilde h^{\infty}_{\D,l,1}(r,\eta)$ and $\tilde h^{\infty}_{\D,l,2}(r,\eta)$ obtained solving the Casimir equation at subleading in large $\D$, 
{ \small
\begin{align}
\label{htildeinfinity1} 
\begin{split}
\tilde h^{\infty}_{\D,l,1}(r,\eta)=&\frac{ 2^{-l} l!}{l \left(r^2-2 \eta  r+1\right) \left(r^2+2 \eta  r+1\right) (h-1)_l} \times \Big \{ 
2 (h-1) C_{l-1}^{(h)}(\eta )
 \times \\
&\Big[\eta   \left(r^4+\left(2-4 \eta ^2\right) r^2+1\right) \left(\Delta(1- r^2) +h \left((h-1) r^2-2\right)+r^2+1\right)\\
&+\Delta _{12}^2 \eta  r^2 \left(r^2-1\right) \left(-2 \eta ^2+r^2+1\right) \\
&+\Delta _{34} r \left(\Delta _{34} \eta  r \left(r^2-1\right) \left(-2 \eta ^2+r^2+1\right)+2 \left(\eta ^2-1\right) \left(r^2+1\right)^2\right) \\
&+\Delta _{12} \eta  \left(-2 \Delta _{34} \eta  \left(r^2-1\right)^2 r+r^2 \left(-\left(8 \eta ^2+r^4-3 r^2-5\right)\right)+1\right)
\Big]\\
&-l \left(r^2+1\right) (2 h+l-2) \left(r^2+2 \eta  r+1\right) \left(r^2-2 \eta  r+1\right) C_l^{(h-1)}(\eta )
\Big\} \ , \\
\end{split} 
\\
\label{htildeinfinity2} 
\begin{split}
\tilde h^{\infty}_{\D,l,2}(r,\eta)=&\frac{ 2^{-l}\left(r^2+1\right) l!}{l \left(r^2-2 \eta  r+1\right) \left(r^2+2 \eta  r+1\right) (h-1)_l} \times \Big\{ -\frac{(h-1)}{r} C_{l-1}^{(h)}(\eta ) \times \\
&\Big[\left(r^4+\left(2-4 \eta ^2\right) r^2+1\right) \left(\Delta(1-  r^2 )+h \left((h-1) r^2-2\right)+r^2+1\right) \\
&+\Delta _{34}^2 r^2 \left(r^2-1\right) \left(-2 \eta ^2+r^2+1\right)\\
&+\Delta _{12}^2 \left(r^2-1\right) r^2 \left(-2 \eta ^2+r^2+1\right)\\
&+\Delta _{12} \left( r^2  \left(-\left(r^4-4 \eta ^2 \left(r^2-1\right)+r^2-8 \eta ^3 r+8 \eta  r-1\right)\right)+1 \right) \\
&-2 \Delta _{34} r \left(2 \left(\eta ^2-1\right) \left(r^3+r\right)+\Delta _{12} \eta  \left(r^2-1\right)^2\right)\Big] \\
&-l (2 h+l-2) \left(r^2-2 \eta  r+1\right) \left(r^2+2 \eta  r+1\right) C_l^{(h-1)}(\eta )\Big\} \ .\\
\end{split}
\end{align}
}
Moreover the condition for the conservation of the blocks can be written as 
\be
\textfrak{D}_{\D}\left[h_{\D l,s}(r,\eta)\right]=0  \ , \nonumber
\ee
 where
\begin{align} 
& \label{frakD}
\textfrak{D}_{\D}\left[h_{\D l,s}(r,\eta)\right]\equiv \\
& \nonumber
+\left(r-r^3\right)  \partial_r \tilde h_{\D l,1}(r,\eta)
+2 \left(\eta ^2-1\right) r  \partial_\eta \tilde h_{\D l,1}(r,\eta) \\
& \nonumber
+\eta  r \left(r^2-1\right)   \partial_r \tilde h_{\D l,2}(r,\eta)
+\left(\eta ^2-1\right) \left(r^2+1\right)  \partial_\eta \tilde h_{\D l,2}(r,\eta) \\
&\nonumber
+\left[\Delta (1- r^2) +\frac{4 r \left(\Delta _{43}+2 h-1\right) (\eta +r (\eta  r-2))-\Delta _2 \left(r^4+\left(4 \eta ^2-6\right) r^2+1\right)}{r^2-2 \eta  r+1}\right]  \tilde h_{\D l,1}(r,\eta) \\
& \nonumber
+\left((2 h-1) (\eta +r (\eta  r-2))+\Delta _2 (\eta +r (\eta  r+2))+\Delta  \eta  \left(r^2-1\right)+4 r \Delta _{34} r\right)  \tilde h_{\D l,2}(r,\eta) \ . 
\end{align}

\section{Further Discussions on Jantzen\rq{}s Criterion} \label{App:Jantzen}
In this appendix we discuss the Jantzen\rq{}s criterion in more detail (see section \ref{Section:PVM}).

First notice that for any weight $\l$ that belongs to a wall $\Omega_\g$, its Weyl reflection $s_\b \cdot \l$ is contained in the Weyl reflected wall $\Omega_{s_\b(\g)}$, where $s_\b(\g)$ is the undotted Weyl action
\be
s_{\b}(\g)\equiv \g- 2\frac{ \langle \g, \b  \rangle}{\langle \b, \b  \rangle} \b \ .
\ee
This simple observation is useful to obtain a very efficient way to see when (\ref{eq:Jantzen}) holds. First we check if $\l $ belongs to some wall $\Omega_\g$ for some root $\g$. As we already explained in section \ref{Section:PVM}, if such a wall does not exist either $\Psi^+_\l$ is empty and $M_{\subalg}(\l)$ is trivially simple, or $\Psi^+_\l$ is not empty and $M_{\subalg}(\l)$ is trivially not simple. Then we apply Weyl transformations $s_\b$ for $\b \in  \Psi^+_\l$ to the wall $\Omega_\g$. This generates a new set of walls $\Omega_{\g_\b}$ with
\be
\Omega_{\g_\b} \equiv \Omega_{s_\b(\g)} \ ,\qquad  \b \in  \Psi^+_\l \ .
\ee
The last step is to find for each $\b \in  \Psi^+_\l$ a Weyl transformation $\omega_\b \in W_\subalg$ with odd length that acts on $\Omega_{\g_\b}$, in such a way the set of walls maps to itself
\be
\{ \Omega_{\omega_\b(\g_\b)} \}_{\b \in \Psi^+_\l } =\{ \Omega_{\g_\b} \}_{\b \in  \Psi^+_\l } \ . \label{EasierJantzen0}
\ee
If this set of $\omega_\b$ exists, then $M_{\subalg}(\l)$ is simple. As we will see there is one more simplification: anytime that $\g_\b \in \Phi_\subalg$ then it is trivial to find a $\omega_\b$ such that  $\g_\b$ is invariant under $\omega_\b$. Thus one needs to find a transformation $\omega_\b$ to fulfill (\ref{EasierJantzen0}) only  for the roots  $\g_\b \notin \Phi_\subalg$.

The full criterion can be written as follows. When $\l \in \Omega_\g$, the parabolic Verma module $M_\subalg(\l)$ is simple if for any $\b \in \Psi^+_\l$, it exists $\omega_\b \in W_\subalg$ with odd length such that
\be  \label{EasierJantzen}
\{ \omega_\b(s_\b(\g)) \}_{\b \in \Psi^+_\l } =\{ \pm s_\b(\g) \}_{\b \in  \Psi^+_\l } \ .
\ee
Where  $\pm$ means that the two sets of roots have to be equal up to signs since $\Omega_\g=\Omega_{- \g}$.

\subsection{Example: \texorpdfstring{$\so(5)$}{so(5)}}
Following the algorithm (\ref{EasierJantzen}) we first find that $ \Psi^+_\l=\{(1,1),(1,0)\}$. Then we obtain the roots that parametrize the walls reflected  by the two Weyl reflections $s_{\b}$ with $\b \in \Psi^+_\l$
\be
\begin{array}{l}
s_{(1,1)}\left((1,-1)\right)= (1,-1) \ , \\
s_{(1,0)}\left((1,-1)\right)= (-1,-1)  \ .
\end{array} 
\ee
As a last step we check that the walls $\Omega_{(1,-1)}$ and $\Omega_{(-1,-1)}$ can be reflected into each other by a Weyl transformation of $W_\subalg$ with odd length. In fact  $s_{(0,1)}((1,-1))=-(-1,-1)$. Again the minus sign is not important since $\Omega_\g=\Omega_{-\g}$.

\subsection{Odd Spacetime Dimension}
We first consider the case in which $\l$ belongs to the wall of the Weyl chamber $\Omega_{\a^{++}_{0k}}$, given by  
\be
\Delta=2N+1+l_k-k \qquad (k=1,\dots,N) \ .
\ee
 We obtain $\Psi_\lambda^+=\{\alpha_{01}^{++},
\alpha_{02}^{++},\dots,\alpha_{0,k-1}^{++}\}$.
 Notice that  $s_{\a_{0j}^{++}}(\a^{++}_{0k})=\a^{-+}_{jk} $ for any $1\le j< k \le N$. Therefore each reflected wall is parametrized by a root $\a^{-+}_{jk}$, which can be Weyl reflected to itself using a Weyl symmetry $s_{\a^{++}_{jk}}\in W_\subalg$ with odd length. Therefore $M_{\subalg}(\l)$ is simple. 
The other special case is when $\l$ belongs to the wall $\Omega_{\a^{-+}_{0k}}$,  
\be
\Delta=k-l_k \qquad  (k=1,2,\dots,N) \ .
\ee
 In this case $
\Psi_\lambda^+=\{\a_0^+,\alpha_{01}^{++},
\alpha_{02}^{++},\dots,\alpha_{0N}^{++},
\alpha_{0,k+1}^{+-},\dots,\alpha_{0N}^{+-}\} $. When we Weyl-reflect the wall $\Omega_{\a^{-+}_{0k}}$ with $s_{\b}$ for all the roots $\b \in \Psi_\lambda^+$, we obtain walls parametrized by the following roots:
\begin{align}
s_{ \alpha_{0j}^{+-}}(\a^{-+}_{0k})&=\a^{-+}_{jk}  \qquad (j\neq k) \ ,\label{TrasfRoots1}\\ 
s_{ \alpha_{0j}^{++}}(\a^{-+}_{0k})&=\a^{++}_{jk}   \qquad (j\neq k) \ , \label{TrasfRoots2} \\
s_{\a_{0k}^{++}}(\a^{-+}_{0k})&=\a^{-+}_{0k} \ , \label{TrasfRoots3} \\
s_{\a_0^+}(\a^{-+}_{0k})&=\a^{++}_{0k} \ . \label{TrasfRoots4}
\end{align}
Notice that the cases (\ref{TrasfRoots1}) and (\ref{TrasfRoots2}) are trivial since the transformed roots live in $\Phi_\subalg$, so we can map each one of them to itself using a transformation in $W_\subalg$ with odd length, namely $s_{\a_{jk}^{++}}(\a^{+-}_{jk})=\a^{+-}_{jk}$ and $s_{\a_{jk}^{+-}}(\a^{++}_{jk})=\a^{++}_{jk}$. The two remaining cases  \eqref{TrasfRoots3} and  \eqref{TrasfRoots4} are less trivial because the roots $\a^{-+}_{0k}$ and $\a^{--}_{0k}$ do not live in $\Phi_\subalg$. This means that we cannot map each one to itself, but we can still map one into the other  $s_{\a_{k}^+}( \a^{-+}_{0k})=- \a^{++}_{0k}$ (up to a sign, which is irrelevant since $\Omega_{\a}=\Omega_{-\a}$). Therefore $M_{\subalg}(\l)$ is simple.

\subsection{Even Spacetime Dimension}
The even dimensional case looks much less straightforward then the odd one. This is both because the roots $\a^{+}_{k}$ are absent and because $l_N$ can now be also negative. The result can be sketched as follows
\begin{align}
\l \in\Omega_{\a^{++}_{0k}}&=\{\l: \Delta=2N+l_k-k \}  \Longrightarrow \quad  \begin{array}{|c|cc|}
\hline
& \quad l_N \ge 0 \quad &\quad  l_N<0 \quad \\
\hline
k=1,\dots,N-1 & \mbox{simple} & \mbox{simple}\\
k=N& \mbox{simple} & \mbox{non simple}\\
\hline
\end{array} \\
\nonumber \\
\l \in \Omega_{\a^{+-}_{0k}}&=\{\l: \Delta=k-l_k \}  \qquad \,\,\,\, \Longrightarrow  \quad  \begin{array}{|c|cc|}
\hline
& \quad l_N > 0 \quad &\quad  l_N \le 0 \quad \\
\hline
k=1,\dots,N-1 & \mbox{non simple} & \mbox{non simple}\\
k=N& \mbox{non simple} & \mbox{simple}\\
\hline
\end{array}  
\end{align}
Let us first consider the case in which $\l$ belongs to the wall $\Omega_{\a^{++}_{0k}}$ with $k= 1,\dots,N$. We find that 
\be
\Psi_\lambda^+=\left\{
\begin{array}{l c}
\{\alpha_{01}^{++},
\alpha_{02}^{++},\dots,\alpha_{0,N-1}^{++},\alpha_{0,N}^{+-}\} \qquad & \mbox{if } l_N<0 \mbox{ and } k=N \ ,\\
\{\alpha_{01}^{++},
\alpha_{02}^{++},\dots,\alpha_{0,k-1}^{++}\} & \mbox{otherwise} \ .\\
\end{array}\right. 
\ee
As we already discussed in the odd dimensional case, all the roots $\alpha_{0j}^{++}$ for $j=1,\dots k$ trivially satisfy the Jantzen\rq{}s criterion since $s_{\alpha_{0j}^{++}}(\a^{++}_{0k}) \in \Phi_\subalg$. Thus when we only have roots $\alpha_{0j}^{++}$ with for $j=1,\dots k$, the module $M_{\subalg}(\l)$ is simple. When $l_N<0$  and $k=N$, we also have the root $\alpha_{0N}^{+-}$ which reflect the wall as $s_{\alpha_{0N}^{+-}}(\alpha_{0N}^{++})=\alpha_{0N}^{++}$. Clearly there is no odd transformation in $W_\subalg$ which maps this wall to itself. We can then conclude that the module is not simple.  
The other special case is when $\l$ belongs to the wall $\Omega_{\a^{-+}_{0k}}$.
We obtain 
\be
\Psi_\lambda^+=\left\{
\begin{array}{l c}
\{\alpha_{01}^{++},
\alpha_{02}^{++},\dots,\alpha_{0N-1}^{++}\} \qquad & \mbox{if } l_N \le 0 \mbox{ and } k=N  \ ,\\
\{\alpha_{01}^{++},
\alpha_{02}^{++},\dots,\alpha_{0N}^{++},
\alpha_{0,k+1}^{+-},\dots,\alpha_{0N}^{+-}\} & \mbox{otherwise} \ .\\
\end{array}\right.
\ee
 When we Weyl-reflect the wall $\Omega_{\a^{-+}_{0k}}$ with $s_{\b}$ for all the roots $\b \in \Psi_\lambda^+$, we obtain walls parametrized by the following roots
\begin{align}
s_{ \alpha_{0j}^{+-}}(\a^{-+}_{0k})&=\a^{-+}_{jk} \qquad ( j=1,\dots N, j\neq k) \ ,\label{TrasfRoots1odd}\\ 
s_{ \alpha_{0j}^{++}}(\a^{-+}_{0k})&=\a^{++}_{jk} \qquad (j=1,\dots N, j\neq k) \ , \label{TrasfRoots2odd} \\
s_{\a_{0k}^{++}}(\a^{-+}_{0k})&=\a^{-+}_{0k} \ . \label{TrasfRoots3odd} 
\end{align}
Again (\ref{TrasfRoots1odd}) and (\ref{TrasfRoots2odd}) are trivial since they live in $\Phi_\subalg$. The only root that does not belong to $\Phi_\subalg$ is $\a^{-+}_{0k} $. Since $\a^{-+}_{0k}$ it is alone it does not belong to $\Phi_\subalg$, we conclude that $M_{\subalg}(\l)$ is not simple. It is crucial that for $l_N \le 0$ and $\l \in \Omega_{\a_{0N}^{-+}}$, the root $\a^{++}_{0N}$ is not in $ \Psi_\lambda^+$. Therefore in this case $M_{\subalg}(\l)$ is  simple.

\section{The Kazhdan-Lusztig Conjecture}
\label{app.KLconjecture}

In this appendix let us be more explicit about the KL polynomial $P_{w,x}(q)$ in  \eqref{m_P_KL}.

\subsection{Odd Spacetime Dimension}

Let us first consider the odd-dimensional case: $\so(2N+3)$. Then $W_{\mathfrak{p}}\backslash W$ contains $2N+2$ elements $w_1, \dots, w_{2N+2}$, and the Bruhat ordering among them is shown in figure \ref{Bruhat_ordering}, namely $w_i \le w_j$ when $i\le j$.
 
\begin{figure}[t]
\graphicspath{{Fig/}}  
\def\svgwidth{12 cm}\centering
 \input{Bruhat.pdf_tex}
\caption{Bruhat ordering for elements of $W_{\mathfrak{p}}\backslash W$, for 
$B_{N+1}=\so(2N+3)$ (above) and $D_{N+1}=\so(2N+2)$ (below).}
\label{Bruhat_ordering}
\end{figure} 

In this case, the value of the KL polynomial at $q=1$ is known to be \cite[section 5]{Boe}
\begin{align}
P_{w_i, w_j}(q=1)=
\begin{cases}
1 & (w_i \ge w_j)   \ , \\
0 & (\textrm{otherwise}) \ . 
\end{cases} 
\end{align}
This means
\begin{align} \label{KLOddDim}
\text{ch}\, L(w_i \cdot \lambda)
= \sum_{j\le i} (-1)^{j-i} \textrm{ch}\, M_{\subalg}(w_j \cdot \lambda)  \ ,
\end{align}
and hence by inverting the relations we obtain
\begin{align}
&\text{ch}\, M_{\subalg}(w_1 \cdot \lambda)=\text{ch}\, L(w_1 \cdot \lambda)  \ ,\\
&\text{ch}\, M_{\subalg}(w_i \cdot \lambda)=\text{ch}\, L( w_i \cdot \lambda)+\text{ch}\, L( w_{i-1} \cdot \lambda) \qquad (i\ge 2) \ .
\end{align}
Namely each generalized Verma module contains either one or two irreducible components,
and we do not need to worry about the null descendants within null descendants.
This explains the absence of higher order poles in the corresponding conformal blocks.

\begin{figure}[t]
\graphicspath{{Fig/}}  
\def\svgwidth{7 cm}\centering
 \input{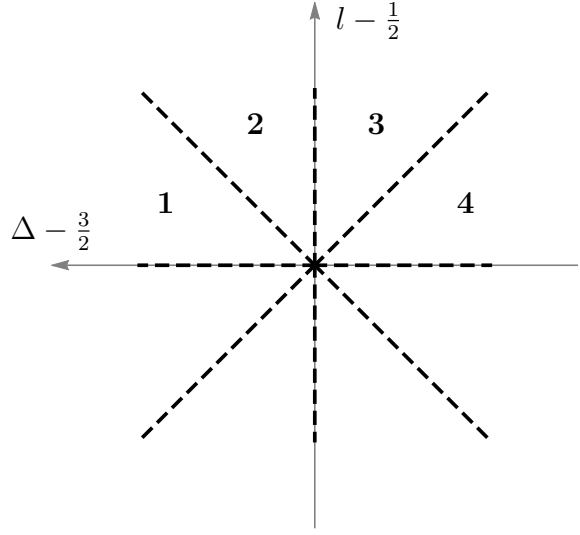}
\caption{Bruhat ordering for $\so(5)$.}
\label{Bruhat_ordering_so(5)}
\end{figure}

Let us take the example of $\so(5)$. The Bruhat ordering is shown in picture \ref{Bruhat_ordering_so(5)}. Notice that the chambers below the origin do not have any label because there is no element in $W_\subalg \backslash W$ which can map the upper half plane to the lower one (since the root $(0,1)$ is in $\Phi_\subalg$).  Defining $M_{\subalg \, n}\equiv M_{\subalg}(w_n \cdot \l)$ and $L_{n} \equiv L(w_n \cdot \l)$ and applying formula (\ref{KLOddDim}) we obtain
\begin{align}
\text{ch}\, M_{\subalg \, 1}&=\text{ch}\, L_{1} \ , \\
 \text{ch}\, M_{\subalg\, 2}&=\text{ch}\, L_{2}+\text{ch}\, L_{1} \ ,\\
\text{ch}\, M_{\subalg \, 3}&=\text{ch}\, L_{3}+\text{ch}\, L_{2} \ ,\\
\text{ch}\, M_{\subalg \, 4}&=\text{ch}\, L_{4}+\text{ch}\, L_{3}  \ ,
\end{align}
which matches formulas (\ref{KLso(5):C}-\ref{KLso(5):B}) in the main text.

\subsection{Even Spacetime Dimension}

The situation is different for even dimensions: $\so(2N+2)$.
In this case, $W_{\mathfrak{p}}\backslash W$ contains $2N+2$ elements, and the Bruhat ordering among them is a bit more involved, as shown in figure \ref{Bruhat_ordering}: $w_i<w_j$ for $i<j$, except that $w_{N+1}$ and $w_{N+2}$ are uncomparable.
The $q=1$ value of the KL polynomial is known to be
\begin{align}
P_{w_i, w_j}(q=1)=
\begin{cases}
1 & (w_i \le  w_j ) \ ,\\
2 & (n+2 \le j \le 2n-1 , 1\le i \le 2n-j ) \ ,\\
0 & (\textrm{otherwise}) \ .
\label{even_KL}
\end{cases}
\end{align}

\begin{figure}[t!]
\graphicspath{{Fig/}}  
\def\svgwidth{12 cm}\centering
 \input{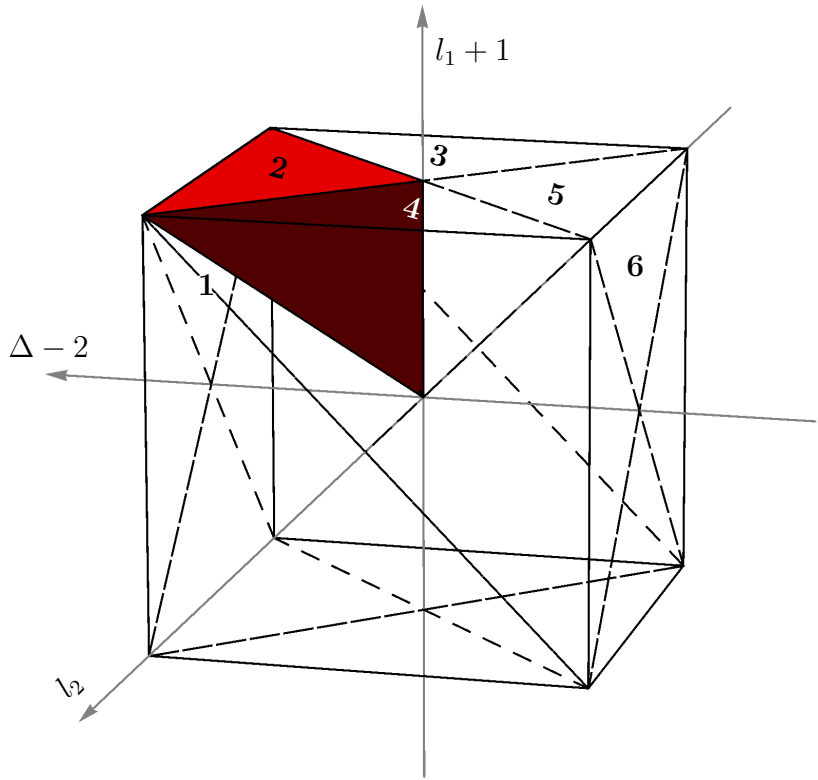}
\caption{\label{WeylSO6} Bruhat ordering of $\so(6)$. The Weyl chambers are the cones spanned by the lines originating from the center of the cube and passing through the three points of the triangles in the faces. We labeled them with numbers in the faces (for example we colored in red the Weyl chamber number $2$). } 
\end{figure}

For example, for $\so(6)$ the Bruhat ordering of the Weyl chambers is shown in figure \ref{WeylSO6}. Again we only labeled the Weyl chambers of the parabolic module. We also show in red the shape of a Weyl chamber, which now are three dimensional objects.
Applying formula \ref{even_KL} we get
\begin{align}
\begin{split}
\text{ch}\, L_{1}&=\text{ch}\, M_{\subalg \, 1} \ ,\\
\text{ch}\, L_{{2}}&=\text{ch}\, M_{\subalg \, 2}-\text{ch}\, M_{\subalg \, 1}\ ,\\
\text{ch}\, L_{{3}}&=\text{ch}\, M_{\subalg \, 3}-\text{ch}\, M_{\subalg \, 2}+\text{ch}\, M_{\subalg \, 1}\ ,\\
\text{ch}\, L_{{4}}&=\text{ch}\, M_{\subalg \, 4}-\text{ch}\, M_{\subalg \, 2}+\text{ch}\, M_{\subalg \, 1}\ ,\\
\text{ch}\, L_{{5}}&=\text{ch}\, M_{\subalg \, 5}-\text{ch}\, M_{\subalg \, 4}-\text{ch}\, M_{\subalg \, 3}+\text{ch}\, M_{\subalg \, 2}-2\, \text{ch}\, M_{\subalg \, 1}\ ,\\
\text{ch} L_{{6}}&=\text{ch}\, M_{\subalg \, 6}-\text{ch}\, M_{\subalg \, 5}+\text{ch}\, M_{\subalg \, 4}+\text{ch}\, M_{\subalg \, 3}-\text{ch}\, M_{\subalg \, 2}+\text{ch}\, M_{\subalg \, 1}\ .
\end{split}
\end{align}
and hence
\begin{align}
\begin{split}
\text{ch}\, M_{\subalg \, 1}&=\text{ch}\, L_1 \ ,\\
\text{ch}\, M_{\subalg \, {2}}&=\text{ch}\, L_{2}+\text{ch}\, L_{1}\ ,\\
\text{ch}\, M_{\subalg \, {3}}&=\text{ch}\, L_{3}+\text{ch}\, L_{2}\ ,\\
\text{ch}\, M_{\subalg \, {4}}&=\text{ch}\, L_{4}+\text{ch}\, L_{2}\ ,\\
\text{ch}\, M_{\subalg \, {5}}&=\text{ch}\, L_{5}+\text{ch}\, L_{4}+\text{ch}\, L_{3}+\text{ch}\, L_{2}+\text{ch}\, L_{1}\ ,\\
\text{ch} M_{\subalg \, {6}}&=\text{ch}\, L_{6}+\text{ch}\, L_{5}+\text{ch}\, L_{1}\ .
\end{split}
\end{align}
In particular, the parabolic Verma modules contain in general more than two irreducible components.

\clearpage

\bibliographystyle{./utphys}
\bibliography{./mybib}

\end{document}